\documentclass{article}

\usepackage{PRIMEarxiv}

\usepackage[utf8]{inputenc} 
\usepackage[T1]{fontenc}    
\usepackage{hyperref}       
\usepackage{url}            
\usepackage{booktabs}       
\usepackage{amsmath,amsfonts}
\usepackage{nicefrac}       
\usepackage{microtype}      
\usepackage{fancyhdr}       
\usepackage{graphicx}       
\usepackage{algorithmic}
\usepackage[]{algorithm2e}
\usepackage{array}
\usepackage[caption=false,font=normalsize,labelfont=sf,textfont=sf]{subfig}
\usepackage{textcomp}
\usepackage{pifont}
\usepackage{float}

\title{Towards Practical Emotion Recognition: An Unsupervised Source-Free Approach for EEG Domain Adaptation}

\author{
  Md Niaz Imtiaz \\
  Department of Electrical, Computer and Biomedical Engineering \\
  Toronto Metropolitan University \\
  Toronto, ON M5B 2K3, Canada \\
  \texttt{niaz.imtiaz@torontomu.ca} \\
  \And
  Naimul Khan \\
  Department of Electrical, Computer and Biomedical Engineering \\
  Toronto Metropolitan University \\
  Toronto, ON M5B 2K3, Canada \\
  \texttt{n77khan@torontomu.ca} \\
}

\begin{document}
\maketitle

\begin{abstract}
Emotion recognition is crucial for advancing mental health, healthcare, and technologies like brain-computer interfaces (BCIs). However, EEG-based emotion recognition models face challenges in cross-domain applications due to the high cost of labeled data and variations in EEG signals from individual differences and recording conditions. Unsupervised domain adaptation methods typically require access to source domain data, which may not always be feasible in real-world scenarios due to privacy and computational constraints. Source-free unsupervised domain adaptation (SF-UDA) has recently emerged as a solution, enabling target domain adaptation without source data, but its application in emotion recognition remains unexplored. We propose a novel SF-UDA approach for EEG-based emotion classification across domains, introducing a multi-stage framework that enhances model adaptability without requiring source data. Our approach incorporates Dual-Loss Adaptive Regularization (DLAR) to minimize prediction discrepancies on confident samples and align predictions with expected pseudo-labels. Additionally, we introduce Localized Consistency Learning (LCL), which enforces local consistency by promoting similar predictions from reliable neighbors. These techniques together address domain shift and reduce the impact of noisy pseudo-labels, a key challenge in traditional SF-UDA models. Experiments on three widely used datasets, DEAP, SEED, and DREAMER, demonstrate the effectiveness of our method. Our approach significantly outperforms state-of-the-art methods, achieving 65.84\% and 58.87\% accuracy on SEED and DREAMER, respectively, when trained on DEAP, and 58.99\% and 67.08\% accuracy on DEAP and DREAMER, respectively, when trained on SEED. It excels at detecting both positive and negative emotions, making it well-suited for practical emotion recognition applications. Code available at \url{https://github.com/RyersonMultimediaLab/EmotionRecognitionSF-UDA}
\end{abstract}

\section{Introduction}
\label{introduction}
Emotion, a fundamental mental phenomenon, is deeply connected to human cognition and behavior, shaping daily life experiences. Recognizing emotions is a critical focus in psychology, medicine, and engineering. In psychology, it provides a quantitative foundation for understanding emotion-related behaviors. In medicine, it aids in the diagnosis and treatment of conditions such as autism spectrum disorder (ASD) and depression and helps track patient recovery \cite{rashidan2021technology,joshi2013multimodal}. In engineering, emotion recognition advances affective brain-computer interfaces (aBCIs), enabling better integration of emotional states into technology.

Physiological signals, such as electroencephalogram (EEG), offer a more objective approach to emotion recognition than non-physiological signals like facial expressions, gestures, and voice, which can be influenced by environmental or social factors \cite{bota2019review}. EEG measures brain activity through electrical signals, providing high reliability and precision, while also being portable and cost-effective.

Advancements in deep learning have greatly improved EEG decoding performance through various neural network models \cite{ramzan2023fused,phadikar2023unsupervised,yousefi2023enhancing}. However, these models generally perform well on test data with a distribution similar to the training data, but their performance significantly declines when applied to a new domain with a different data distribution, a phenomenon known as domain shift. The considerable variations in EEG signals across individuals challenge the effectiveness of cross-domain emotion recognition models. Additionally, deep neural networks typically require a vast quantity of labeled data for training. Collecting and annotating EEG signals is often a labor-intensive and costly process.

Unsupervised domain adaptation (UDA) is gaining attention to address these challenges, aiming to enhance model performance on the target domain by utilizing information from the source domain. Recent research on domain adaptation for classifying EEG emotions has largely focused on cross-subject and cross-session scenarios within the same dataset \cite{10412628,8567966,she2023cross,huang2022generator,li2022dynamic}. However, less attention has been given to the cross-dataset scenario, which presents even greater challenges due to variations in EEG signals not only from individual differences but also from differences in EEG recording settings and stimuli.

Another limitation of traditional unsupervised domain adaptation methods is that they require access to the source domain data while adapting to the target domain. However, in many real-world applications, accessing the source data of a trained source model is not feasible. For example, when deploying domain adaptation algorithms on mobile devices with constrained computational resources, keeping source domain data on the device is impractical. Additionally, to preserve the privacy of source subjects, their data may be inaccessible during domain adaptation. This is particularly true for healthcare applications, where data sharing is complicated due to ethical concerns. With the lack of source samples, traditional distribution matching techniques become infeasible.

\begin{figure}[!tb]
     \centering
    \includegraphics[width=\linewidth]{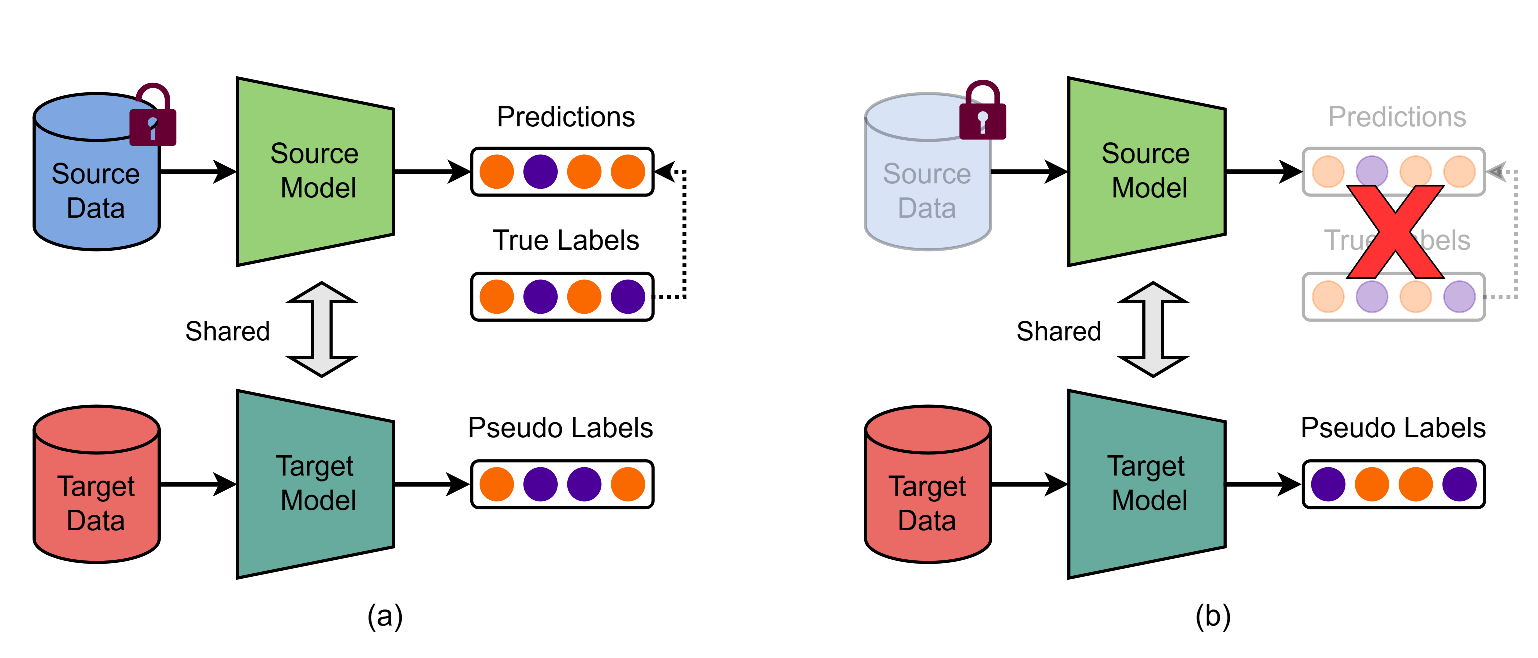}
    \caption{Illustration of domain adaptation strategies: (a) Traditional unsupervised domain adaptation (UDA); (b) Source-free unsupervised domain adaptation (SF-UDA). Unlike traditional UDA, SF-UDA transfers knowledge without accessing source data, thereby preserving data privacy.}
    \label{fig1}
\end{figure}

Source-free unsupervised domain adaptation (SF-UDA) focuses on transferring knowledge from a source domain to a target domain without requiring access to the source dataset, thus inherently protecting individual privacy \cite{liang2020we}. Unlike UDA, where labeled source domain data, the trained model, and unlabeled target data are available, SF-UDA relies solely on the trained model and unlabeled target data, with no access to source domain data (Fig. \ref{fig1}). SF-UDA faces several challenges. Biases in the source data during pre-training can cause the model to focus on domain-specific features rather than learning fundamental patterns. Additionally, without access to source data, target adaptation often suffers due to the generation of noisy pseudo-labels for the target data.

To address the challenges outlined above, we propose a novel source-free unsupervised domain adaptation (SF-UDA) method for EEG-based emotion classification. SF-UDA has primarily been applied to image classification tasks \cite{du2024generation,ahmed2023ssda,lee2023feature,li2022divergence,tian2021vdm,kumar2023conmix}, with only a few studies exploring its use for EEG signal classification, such as seizure and sleep stage classification \cite{zhao2023source,ragab2023source,zhao2024source}. To the best of our knowledge, this is the first application of SF-UDA to EEG-based emotion recognition. Our method comprises four stages: pre-training, computation, target adaptation, and inference. The pre-training and inference stages, as well as the network architecture, a feature extractor followed by two parallel classifiers, are adopted from our previous work \cite{imtiaz2025enhanced}.

During the pre-training stage, the model is trained on labeled source samples to learn essential features for emotion recognition. Once pre-training is complete, the source domain data is no longer accessible for the subsequent stages. The computation stage calculates cluster centroids and classifier discrepancies using the pre-trained model's predictions on the target domain data. These calculations lay the foundation for target adaptation.

The target adaptation stage involves two training steps using unlabeled target data. In the first step, \textit{Dual-Loss Adaptive Regularization (DLAR)}, the model is trained to minimize classifier discrepancies on confident samples while aligning its predictions with the calculated pseudo-labels. This ensures the model produces consistent and reliable predictions. The second step, \textit{Localized Consistency Learning (LCL)}, enforces local consistency by encouraging reliable neighbors to produce similar predictions. To mitigate the influence of unreliable neighbors, we introduce a mechanism for identifying reliable neighbors based on both feature values and softmax probabilities. In the inference stage, we apply our recently proposed Prediction Confidence-aware Test-Time Augmentation (PC-TTA) \cite{imtiaz2025enhanced}, which selectively augments test samples based on predictive confidence, improving model prediction performance while minimizing computational costs.

Our proposed method is evaluated on three widely used emotion recognition datasets: DEAP \cite{koelstra2011deap}, SEED \cite{zheng2015investigating}, and DREAMER \cite{katsigiannis2017dreamer}. To assess its performance in cross-dataset settings, where distribution discrepancies are more pronounced due to differences in subject demographics and recording conditions, we conduct experiments by training the model on one dataset and evaluating it on the other. Since no SF-UDA methods currently exist for EEG emotion recognition, we compare our approach with four recent, open-source SF-UDA methods \cite{ragab2023source,zhao2023source,ahmed2023ssda,du2024generation} that are recognized for their high performance. This comparison is carried out by evaluating all methods on the same datasets used in our experiments. Our method outperforms all others by a significant margin, achieving accuracies of 65.84\% (DEAP $\rightarrow$ SEED), 58.99\% (SEED $\rightarrow$ DEAP), 58.87\% (DEAP $\rightarrow$ DREAMER), and 67.08\% (SEED $\rightarrow$ DREAMER) for binary emotion classification, while also excelling in multi-class settings.

The key contributions of this paper are as follows:

(1) A novel source-free unsupervised domain adaptation (SF-UDA) method is introduced for EEG-based emotion recognition, an area where SF-UDA has not previously been applied.

(2) A robust multi-stage framework is proposed to effectively address the challenges of cross-domain emotion recognition, enhancing the model's ability to adapt across diverse datasets.

(3) A novel technique called Dual-Loss Adaptive Regularization (DLAR) is proposed, which minimizes classifier discrepancies on confident samples and aligns predictions with pseudo-labels, ensuring consistent and reliable performance.

(4) Localized Consistency Learning (LCL) is introduced to enforce local consistency in predictions by encouraging similar outcomes from reliable neighbors.

The structure of the paper is outlined as follows: Section \ref{related_work} provides an overview of the related literature. Section \ref{proposed_method} describes our proposed method in detail, including the model architecture, training and testing processes, and data preprocessing. Section \ref{experiments} presents the experimental setup, discusses the datasets used, and reports the results of the experiments, including both quantitative and qualitative evaluations. The paper concludes in Section \ref{conclusion}.

\section{Related Work}
\label{related_work}
Extensive research has been conducted on classifying EEG signals using machine learning techniques. Commonly used methods include Support Vector Machines \cite{son2021eeg,kumar2021universum}, Decision Trees \cite{chalupnik2022using,xu2022epilepsy}, Logistic Regression \cite{singh2024emotion}, and Linear Discriminant Analysis \cite{panachakel2021automated,fu2020automatic}. However, conventional shallow learning approaches often struggle to capture the complex temporal patterns in EEG signals, limiting their ability to generalize to unseen data. In deep learning, Autoencoders \cite{wang2022multi,phadikar2023unsupervised}, Graph Neural Networks \cite{lin2023eeg}, and Deep Belief Networks \cite{wang2023classification} have been investigated for EEG analysis. Additionally, Convolutional Neural Networks and Long Short-Term Memory networks are widely used for EEG classification tasks \cite{xu2025unsupervised,shi2023enhancing,ramzan2023fused,yousefi2023enhancing}.

While machine learning and deep learning have advanced EEG signal processing \cite{ramzan2023fused,phadikar2023unsupervised,yousefi2023enhancing,shi2023enhancing}, their performance often drops under domain shifts caused by individual, demographic, or recording differences. Unsupervised Domain Adaptation (UDA) addresses this challenge by transferring knowledge from a labeled source domain to an unlabeled target domain \cite{xu2025unsupervised,zhu2025tensorial}. In EEG emotion classification, many UDA approaches focus on feature alignment to enhance generalization by learning a shared feature space or minimizing discrepancies between feature distributions. For classifying emotions across subjects, Jiménez-Guarneros et al. \cite{10412628} proposed a method that aligns subject distributions in the global feature space and fine-tunes the alignment between modalities within each domain's distribution. She et al. \cite{she2023cross} developed a method for selecting a source domain with limited labeled target data, using manifold feature mapping to mitigate data drift and minimum redundancy maximum correlation to improve classifier performance.

Adversarial-based methods, another prominent UDA approach, minimize domain shift through adversarial training to learn domain-invariant features. The feature extractor produces features that are indistinguishable between the source and target domains, while the discriminator distinguishes between them. Huang et al. \cite{huang2022generator} introduced a knowledge-free approach to align source domain features with the target domain through adversarial learning while preserving emotional information using EEG content and emotion-related loss functions. Li et al. \cite{8567966} incorporated hemisphere-specific features through adversarial learning to address domain shifts.

Only a limited number of studies \cite{ni2021domain,gu2022multi,imtiaz2025enhanced} have explored the cross-dataset scenario for EEG emotion recognition, which poses significant challenges. These challenges arise from variations in EEG signals due to individual differences, as well as differences in stimuli and recording settings. Ni et al. \cite{ni2021domain} proposed a domain-adaptive sparse representation classifier that projects source and target samples into a shared subspace while learning a domain-invariant dictionary. Gu et al. \cite{gu2022multi} combined transfer learning with dictionary learning using subspace projection. In our prior work \cite{imtiaz2025enhanced}, we introduced a gradual proximity-guided target selection method to reduce negative transfer and a cost-effective test-time augmentation strategy that applies augmentations only when needed.

Conventional UDA methods rely on accessing source data during adaptation, which is not always feasible due to privacy concerns or computational limitations. To overcome this, researchers have recently turned to source-free domain adaptation (SFDA), which facilitates knowledge transfer using a source-pretrained model and unlabeled target data without requiring access to the original source dataset. SFDA has been extensively explored in the field of computer vision \cite{du2024generation,ahmed2023ssda,lee2023feature,li2022divergence,tian2021vdm,kumar2023conmix}. Based on existing works, SFDA can be categorized into two directions \cite{li2024comprehensive}: data-centric \cite{lee2023feature,li2022divergence,du2024generation,li2020model,tian2021vdm,ding2023proxymix} and model-centric \cite{ahmed2023ssda,liang2020we,kothandaraman2023salad,ding2022source,kumar2023conmix,liu2021source}.

Data-centric methods aim to reconstruct or subdivide the target domain to compensate for the absence of source domain data, facilitating the extension of UDA methods to the SFDA setting. These methods focus on enhancing the target domain's data to refine its representation and improve model generalization. The method proposed by Lee and Lee \cite{lee2023feature} reduces the domain gap by aligning prediction and feature spaces, incorporating data augmentation and consistency objectives, while also addressing epistemic uncertainty to enhance model adaptation. Li et al. \cite{li2022divergence} proposed an approach that designs adversarial examples to attack the training model, leveraging these examples to improve generalization and address the divergence-agnostic learning challenge. In their method, Du et al. \cite{du2024generation} create and expand a pseudo-source domain using target samples to reduce domain discrepancy, utilizing four loss functions.

Model-centric methods assume that the source-pretrained model generalizes well to the target domain due to shared characteristics. Fine-tuning the model by leveraging its predictions on the target data helps bridge the domain gap, improving adaptability through adjustments in the model's structure, parameters, or training strategy. Liang et al. \cite{liang2020we} proposed a method that freezes the source classifier and adapts the feature extractor to the target domain using information maximization and pseudo-labeling. Ahmed et al. \cite{ahmed2023ssda} combined model compression with spectral-norm-based loss penalties to remove malicious channels, restoring accuracy through dynamic knowledge transfer during target domain training. Kothandaraman \cite{kothandaraman2023salad} introduced an approach for adapting a pre-trained network to a target domain with limited annotations and label space shifts, optimizing sample selection and knowledge transfer for task-agnostic, source-free adaptation.

While considerable research on SFDA has been conducted using image data, few studies have explored its application to time series data \cite{ragab2023source,zhao2023source,zhao2024source}. Ragab et al. \cite{ragab2023source} presented a method that captures temporal information through random masking and a temporal imputer, guiding the target model to align with source domain features while preserving temporal consistency. Zhao and Peng \cite{zhao2023source} introduced SS-TrBoosting, a boosting-based SFDA approach for EEG seizure subtype classification, and extended it to U-TrBoosting for unsupervised SFDA.

Despite remarkable progress in deep learning for EEG-based emotion recognition, challenges remain in cross-domain contexts, particularly when significant domain shifts create substantial disparities between source and target domains. The absence of source samples further complicates the situation, making traditional distribution-matching methods ineffective. While source-free domain adaptation offers a potential solution to these issues, its application in emotion recognition has yet to be explored. To address these gaps, we propose a novel source-free unsupervised domain adaptation method tailored for EEG-based emotion classification. Our approach focuses on mitigating domain discrepancies in cross-dataset scenarios, overcoming the scarcity of labeled data, and handling the unavailability of source datasets during target adaptation.

\section{Proposed Method}
\label{proposed_method}
Our proposed approach consists of four stages: pre-training, computation, target adaptation, and inference. The pre-training and inference stages remain consistent with our prior UDA model \cite{imtiaz2025enhanced}. However, we introduce new computation and target adaptation stages to facilitate efficient adaptation to the target domain without relying on data from the source domain.

\subsection{Framework}
\begin{figure}[!tb]
     \centering
    \includegraphics[width=\linewidth]{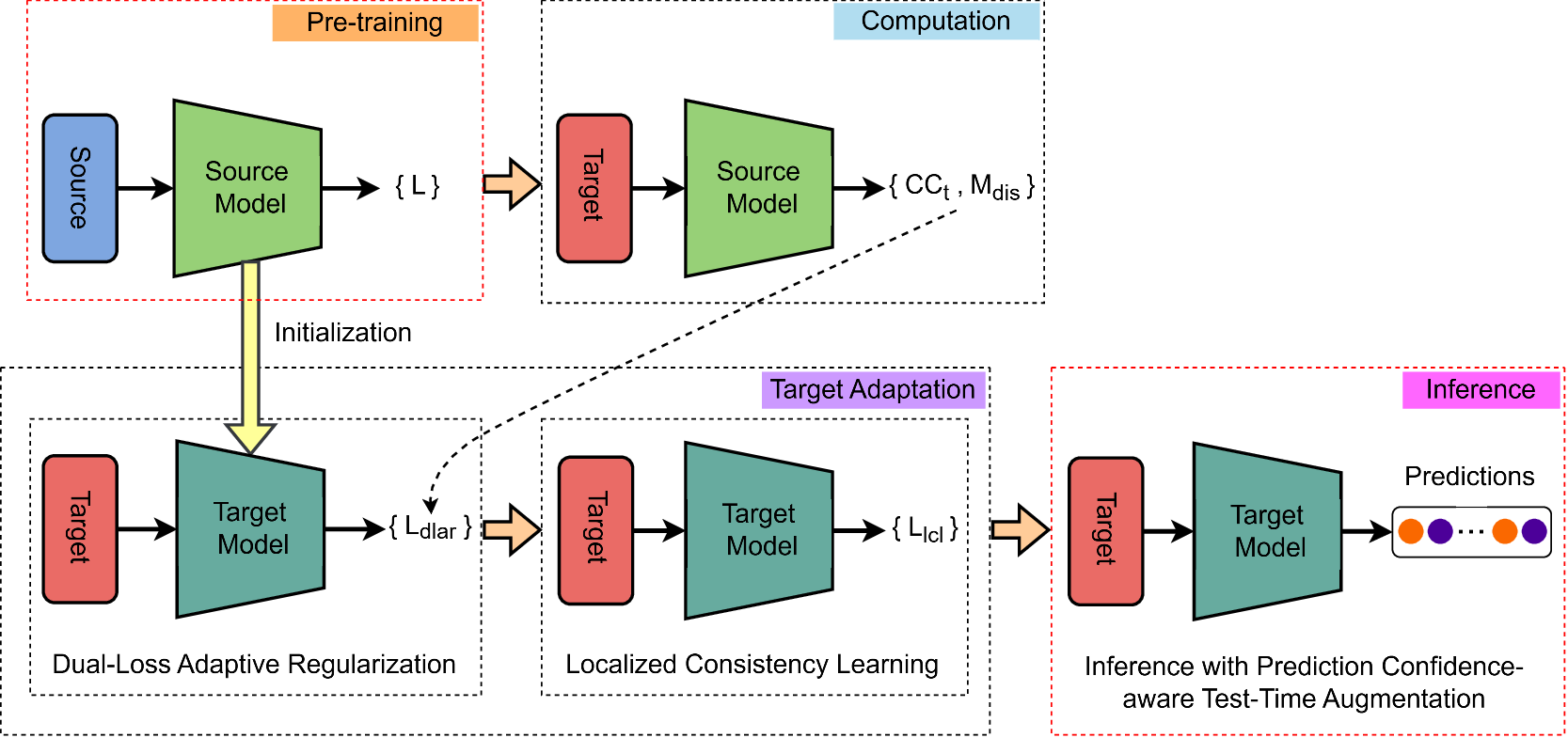}
    \caption{Overview of the proposed method. The stages enclosed in red-bordered boxes are adopted from our recent UDA model \cite{imtiaz2025enhanced}, while the stages enclosed in black-bordered boxes represent the original contributions of this research. The proposed method consists of four stages: pre-training, where features are learned from labeled source data; computation, where target-domain cluster centroids and classifier discrepancies are calculated; target adaptation, where DLAR and LCL enhance prediction consistency and reliability on unlabeled target data; and inference, where model performance on test data is improved using selective augmentations.}
    \label{fig2}
\end{figure}

The framework of the proposed method is presented in Fig. \ref{fig2}. Suppose the source domain consists of $N_s$ labeled samples $X_s$= $\{{x_s^i}\}_{i=1}^{N_s}$ and their corresponding labels $Y_s$= $\{{y_s^i}\}_{i=1}^{N_s}$, while the target domain contains $N_t$ unlabeled samples $X_t$= $\{{x_t^i}\}_{i=1}^{N_t}$. The marginal distributions of the source and target domains are $P_s(X_s)$ and $P_t(X_t)$, respectively, where $P_s(X_s)$ $\neq$ $P_t(X_t)$. This work focuses on solving the source-free domain adaptation challenge, where source data is inaccessible during the adaptation process to maintain data privacy. The aim is to learn a function $f$ that reduces the discrepancy between the marginal distributions $P_s(X_s)$ and $P_t(X_t)$, ensuring accurate predictions for the target samples.

The network architecture follows the structure of our previous model \cite{imtiaz2025enhanced}, which applied cross-dataset EEG emotion recognition in a UDA setting. It consists of a feature extractor ($F$) with two fully connected layers, followed by two parallel classifiers ($C_1$ and $C_2$). Each fully connected layer is followed by batch normalization to standardize features. Previous research \cite{jimenez2023learning,li2019domain} has shown that simple feed-forward networks with fully connected layers perform well for EEG emotion recognition. Given that more complex networks do not improve performance over simpler ones \cite{imtiaz2025enhanced}, we choose the simpler architecture to reduce complexity and mitigate overfitting issues commonly associated with complex models, especially considering the relatively small datasets used in this study. As in our previous models \cite{imtiaz2025enhanced,imtiaz2024cross}, we employ two parallel classifiers after the feature extractor. This dual-classifier setup helps handle situations where a single classifier may fail, even if the feature extractor produces distinct features. Moreover, we utilize the discrepancy between the classifiers to identify confident samples in the target domain during adaptation and assess the need for augmentation in our PC-TTA method. Each classifier has three fully connected layers, and the predicted emotion category is obtained by averaging the outputs of both classifiers.

\subsubsection{Pre-training}
The pre-training stage follows the same approach as in our previous model \cite{imtiaz2025enhanced}. During this phase, the model is trained using labeled source data ($X_s$, $Y_s$), to learn key features for emotion recognition. The loss function is the weighted sum of \textit{classification loss} ($L_{cls}$) and \textit{classifier discrepancy loss} ($L_{dis}$) (Eq. \ref{eq1}). To compute the \textit{classification loss}, we apply group distributionally robust optimization (DRO) \cite{sagawa2019distributionally}, which minimizes the worst-case loss across all data groups, to the weighted cross-entropy loss. The \textit{classifier discrepancy loss} is calculated as the Euclidean distance between the predictions of the two classifiers.
\begin{align}
L= L_{cls} + \alpha  L_{dis}
\label{eq1}
\end{align}
where $\alpha$ is a hyperparameter.

\subsubsection{Computation}

After the pre-training stage, the pre-trained network becomes available, and source domain data is no longer accessible in subsequent stages. During the computation stage, we calculate the essential properties needed for target adaptation: \textit{cluster centroids} ($CC_t$) and \textit{mean classifier discrepancy} ($M_{dis}$), by feeding the unlabeled target data into the pre-trained network. Following Liang et al.'s study \cite{liang2020we}, we compute the cluster centroid for each emotion category ($CC_t^k$) in the target domain:
\begin{align}
CC_t^k= \frac{\Sigma_{x_t\in X_t} \: \hat{p}^k(x_t)F(x_t)}{\Sigma_{x_t\in X_t} \: \hat{p}^k(x_t)}
\label{eq2}
\end{align}
 where $\hat{p}^k(x_t)$ represents the aggregated softmax value for the sample $x_t$ from the two classifiers for the $k^{th}$ emotion category:
\begin{align}
\hat{p}^k(x_t)=avg(\delta^k(C_1(F(x_t))), \delta^k(C_2(F(x_t))))
\label{eq3}
\end{align}
where $\delta$ is the softmax function.

This technique calculates the centroids by integrating the classifier's predictions with the feature extractor's outputs, ensuring that samples are weighted according to the classifier's confidence. In contrast, methods relying solely on feature extractor outputs may produce less reliable centroids, as they overlook class boundaries and classifier confidence, which can lead to issues with ambiguity and domain shifts.

The \textit{mean classifier discrepancy} ($M_{dis}$) is calculated as the average difference between the outputs produced by the two classifiers:
\begin{align}
M_{dis}= \frac{1}{N_t} \Sigma_{i=1}^{N_t} \: D(C_{1,i}, C_{2,i})
\label{eq4}
\end{align}
where $D$ represents the Euclidean distance.

\subsubsection{Target Adaptation}

The target adaptation stage consists of two training steps, using only unlabeled target data with the pre-trained network. In the first step, we introduce the novel Dual-Loss Adaptive Regularization (DLAR) technique to improve the model's consistency and reliability on the unlabeled target data. For each training batch, we generate pseudo-labels by assigning each sample to the nearest cluster, using the cluster centroids computed in the computation stage:
\begin{align}
\hat{y}_t= \underset{k}{arg min} \: D_l(F(x_t),CC_t^k)
\label{eq5}
\end{align}
where $D_l$ represents the Euclidean distance (L2 norm).

We then calculate the \textit{pseudo-label agreement} loss ($L_{plal}$) using cross-entropy between the pseudo-labels and the model's predictions:
\begin{align}
L_{plal}= - \frac{1}{N_{b}} \Sigma_{x_t\in X_{t,b}} \: \Sigma_{k=1}^{K}\: 1_{[k=\hat{y}_t] } \log (\hat{p}^k(x_t))
\label{eq6}
\end{align}
where $X_{t,b}$ represents the target samples in a batch, $N_{b}$ denotes the batch size, and $K$ is the total number of emotion categories.
This loss enables the model to learn in a self-supervised manner, where the pseudo-labels act as surrogates for true labels. By aligning the model's predictions with the underlying cluster distribution, this loss improves the model's ability to adapt to unseen target data.

Additionally, we calculate the \textit{confident classifier discrepancy loss} ($L_{ccdl}$) (Eq. \ref{eq10}) by evaluating the discrepancy between classifiers, focusing on samples with \textit{confident predictions} ($CX_t$). Confident samples are selected based on low prediction entropy and low classifier discrepancy:
\begin{align}
\begin{split}
CX_t= \{x_t\in X_{t,b} \mid D(C_1(F(x_t)), C_2(F(x_t))) < M_{dis} \: \\ and \: H(x_t) < M_H\}
\end{split}
\label{eq7}
\end{align}
 where $H(x_t)$ represents the entropy of the predicted distribution for the sample $x_t$, and $M_H$ is the average entropy calculated over the training batch:
\begin{align}
H(x_t)= - \Sigma_{k=1}^{K}\: \hat{p}^k(x_t) \log(\hat{p}^k(x_t) )
\label{eq8}
\end{align}
\begin{align}
M_H= \frac{1}{N_b} \Sigma_{x_t\in X_{t,b}} \: H(x_t)
\label{eq9}
\end{align}
\begin{align}
L_{ccdl}= \frac{1}{N_c} \Sigma_{x_t\in CX_t} \: D_l (C_1(F(x_t)), C_2(F(x_t)))
\label{eq10}
\end{align}
where $N_c$ denotes the number of samples in $CX_t$.

We select the threshold $M_H$ as it yields the best results in experiments with varying entropy, while $M_{dis}$ is chosen consistently with our previous works \cite{imtiaz2025enhanced,imtiaz2024cross}, where it effectively identifies reliable samples. Together, these thresholds are used to select confident samples based on low prediction entropy and low classifier discrepancy.

This training provides additional supervision through classifier agreement, leading to better generalization on the target data. The final loss function ($L_{dlar}$) is the sum of the \textit{pseudo-label agreement loss} and the \textit{confident classifier discrepancy loss}:
\begin{align}
L_{dlar}= L_{plal} +  L_{ccdl}
\label{eq11}
\end{align}

At the end of each training epoch in DLAR, the cluster centroids are updated by recomputing them using Eq. \ref{eq2} with the updated model.

\begin{figure}[!tb]
     \centering
    \includegraphics[width=0.7\linewidth]{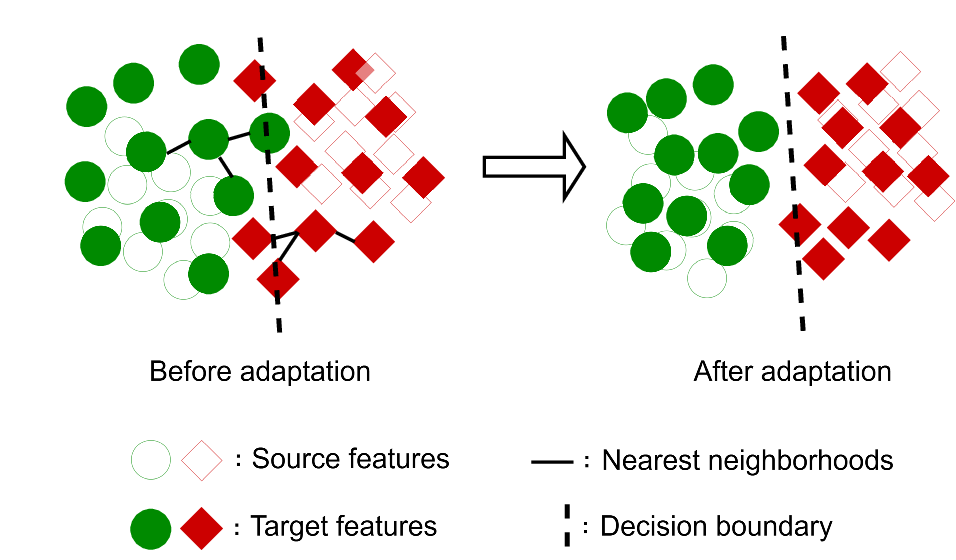}
    \caption{Adaptation using Localized Consistency Learning. This strategy draws reliable neighbors closer in the adapted feature space.}
    \label{fig3}
\end{figure}

In the second step, we introduce the Localized Consistency Learning (LCL) strategy to enhance local consistency by encouraging similar predictions for reliable neighbors. This strategy brings reliable neighbors closer together (Fig. \ref{fig3}). To reduce the influence of unreliable neighbors, we identify trustworthy ones by considering both the feature representations and the softmax probabilities. For each sample $x_i$ in the training batch, we first calculate the $k$-nearest neighbors based on the feature map values ($NN_{feat,i}$) and the softmax outputs ($NN_{cls,i}$):
\begin{align}
NN_{feat,i}= \{x_j \in X_{t,b} \mid x_i \neq x_j \: and  \: x_j \in kNN_{feat}(x_i)\}
\label{eq12}
\end{align}
\begin{align}
NN_{cls,i}= \{x_j \in X_{t,b} \mid x_i \neq x_j \: and  \: x_j \in kNN_{cls}(x_i)\}
\label{eq13}
\end{align}
where $kNN_{feat}(x_i)$ and $kNN_{cls}(x_i)$ refer to the  $k$-nearest neighbors of sample $x_i$, based on the feature map values and softmax outputs, respectively.

Next, we select the common neighbors ($NN_{final,i}$) that appear in both $NN_{feat,i}$ and $NN_{cls,i}$:
\begin{align}
NN_{final,i}= NN_{feat,i} \cap NN_{cls,i}
\label{eq14}
\end{align}
The model is then trained to minimize the prediction discrepancies among these common neighbors using the \textit{localized consistency loss} ($L_{lcl}$) function:
\begin{align}
L_{lcl}= - \frac{1}{N_b} \Sigma_{i=1}^{N_b} \:  \Sigma_{x_j\in NN_{final,i}} \: \log(\hat{p}(x_i) \cdot \hat{p}(x_j))
\label{eq15}
\end{align}

The Dual-Loss Adaptive Regularization (DLAR) technique enhances model consistency and reliability on unlabeled target data while reducing noisy predictions. The Localized Consistency Learning (LCL) strategy further promotes local consistency by encouraging similar predictions among reliable neighbors. It aligns the feature and classifier spaces, ensuring that similar samples receive consistent predictions. Together, these components enable the model to learn robust representations, leading to improved generalization.

\subsubsection{Inference}

During inference, we utilize a Prediction Confidence-aware Test-Time Augmentation (PC-TTA) strategy, as introduced in our previous work \cite{imtiaz2025enhanced}, to improve model performance on target domain test data. TTA typically applies various augmentations to test samples and combines their predictions to enhance robustness. However, this process can be computationally expensive due to the multiple transformations involved. PC-TTA mitigates this issue by selectively applying TTA only to samples with high prediction uncertainty ($u$). Uncertainty is measured using entropy, calculated from the model's softmax probabilities \cite{imtiaz2025enhanced}. If the uncertainty exceeds a defined threshold ($\tau$) and classifier discrepancy is higher than the \textit{mean classifier discrepancy} ($M_{dis}$), TTA is performed; otherwise, the model's initial prediction is accepted. For samples requiring TTA, EEG signal segments are augmented using techniques such as Gaussian noise addition and resampling. Features from these transformations are input into the model to generate predictions. A final label is determined by voting among predictions from the original and augmented samples.

Algorithm \ref{alg} summarizes the complete procedure.

\begin{algorithm}
\footnotesize
\begin{algorithmic}
\REQUIRE
\hfill
\\Source PSD samples and labels: $X_s$, $Y_s$\\
Target PSD samples: $X_t$\\
Feature extractor: $F$, Classifiers: $C_1$, $C_2$\\
Epochs: $Epoch_{dlar}$, $Epoch_{lcl}$\\

\ENSURE
\hfill
\\
\textbf{Pre-training} \\
Train the model using $X_s$ and $Y_s$ with the loss function $L$ (Eq. \ref{eq1})

\textbf{Computation}\\
Compute cluster centroids $CC_t$ (Eq. \ref{eq2}) and mean classifier discrepancy $M_{dis}$ (Eq. \ref{eq4}) from $X_t$ via the pre-trained model\\
\textbf{Target Adaptation}\\
\For{($i$ = 1 to $Epoch_{dlar}$)}{
    Generate pseudo-labels for $X_t$ using Eq. \ref{eq5}\\
    Compute pseudo-label agreement loss ($L_{plal}$) using Eq. \ref{eq6}\\
    Select samples with confident predictions ($CX_t$) using Eq. \ref{eq7}\\
    Compute confident classifier discrepancy loss ($L_{ccdl}$) using Eq. \ref{eq10}\\
    Train the model using $X_t$:\\
 Update the parameters of $F$, $C_1$, and $C_2$ by minimizing the combined loss $L_{dlar}$, comprising $L_{plal}$ and $L_{ccdl}$ (Eq. \ref{eq11})
}
\For{($i$ = 1 to $Epoch_{lcl}$)}{
Identify nearest neighbors for each target sample using Eq. \ref{eq12},\ref{eq13}, and \ref{eq14}\\
Train the model using $X_t$:\\
Update the parameters of $F$, $C_1$, and $C_2$ by minimizing $L_{lcl}$ (Eq. \ref{eq15})
}
\textbf{Inference}\\
\For{(each $x_t$ $\in$ $X_t$)}{
Calculate uncertainty $u$ using entropy from the model's softmax probabilities\\
\If{(uncertainty $u$  $>=$ threshold $\tau$ and \\ \: \:  classifier discrepancy $D(C_1, C_2)$ $>$ $M_{dis}$)}{
Apply augmentation and obtain predictions from the augmented samples\\
Obtain final prediction by majority voting among predictions from the original and augmented samples\\
}
\Else{
    Accept the model's initial prediction for $x_t$ as the final prediction
    }
}

\caption{Steps of the proposed source-free unsupervised domain adaptation approach for cross-dataset EEG emotion recognition.}
\label{alg}
\end{algorithmic}
\end{algorithm}

\subsection{Data Preprocessing and Model Input Construction}
The data preprocessing is carried out using the procedure described in our previous work \cite{imtiaz2025enhanced}. Power Spectral Density (PSD) and Differential Entropy (DE) are commonly used in EEG-based emotion recognition, with prior studies \cite{ni2021domain,li2022eeg} showing that they often outperform other types of EEG features. In this work, we examine both PSD and DE; however, our results demonstrate that PSD yields superior performance. Therefore, we focus on incorporating PSD into our proposed approach.

We extract EEG data from 32 common channels shared by the DEAP and SEED datasets to maintain a consistent input size for the model. Each EEG trial is segmented into 2-second windows with a 1-second overlap. We compute the PSD features for each segment across five frequency bands: delta (1--3 Hz), theta (4--7 Hz), alpha (8--13 Hz), beta (14--30 Hz), and gamma (31--50 Hz). These features are concatenated into a 1-D feature vector, $X \in \mathbb{R}^{n*i}$, where $n$ = 32 denotes the number of channels, and $i$ = 5 corresponds to the five frequency bands. Thus, the 1-D input feature vector has a size of 160 (5$\times$32). In the experiment with the DREAMER dataset, EEG data are extracted from 14 channels to maintain a consistent input size for the model, resulting in an input feature vector of size 70 (5$\times$14). Finally, the features are normalized to the range [-1, 1].

\section{Experiments}
\label{experiments}
\subsection{Experimental Details}
The proposed model is assessed through experiments conducted on the DEAP, SEED, and DREAMER datasets, which are publicly available and extensively used in emotion recognition studies. To validate the model's performance, we perform cross-dataset testing by training it on one dataset and evaluating it on another.

\subsubsection{DEAP Dataset \cite{koelstra2011deap}}
The DEAP dataset includes EEG and peripheral physiological signals from 32 participants who watched 40 one-minute music videos. Each participant completed 40 trials, each lasting 63 seconds, consisting of a 3-second pre-trial phase followed by 60 seconds of video viewing. After each trial, participants rated their arousal, valence, liking, and dominance on nine-point scales. EEG signals were recorded from 32 electrodes at a sampling rate of 512 Hz, then preprocessed by down-sampling to 128 Hz, removing EOG artifacts, and applying a 4--45 Hz bandpass filter.

\subsubsection{SEED Dataset \cite{zheng2015investigating}}
The SEED dataset includes EEG recordings from 15 Chinese participants across three separate sessions conducted on different days. During each session, participants viewed 15 carefully selected Chinese film clips designed to evoke positive, neutral, and negative emotions. EEG data were recorded using a 62-channel electrode system at a sampling rate of 1,000 Hz, then down-sampled to 200 Hz and filtered within the 0--75 Hz frequency range.

\subsubsection{DREAMER Dataset \cite{katsigiannis2017dreamer}}
The DREAMER dataset includes EEG data from 23 participants, each of whom watched 18 film clips ranging from 65 to 393 seconds. Participants rated valence, arousal, and dominance on an integer scale from 1 to 5. EEG signals were recorded from 14 electrodes at a sampling rate of 128 Hz.

For our experiments, we maintain the same data preparation process as in our previous work \cite{imtiaz2025enhanced}. We use data from all subjects and trials in the SEED, DEAP, and DREAMER datasets. For DEAP, we exclude the first 3 seconds of each trial, which are pre-trial data. SEED and DREAMER signals are filtered with a bandpass filter (0.3 Hz to 50 Hz) to remove noise \cite{li2019domain,li2022dynamic,imtiaz2025enhanced}, while DEAP signals, already filtered between 4 Hz and 45 Hz, undergo no additional filtering. This study primarily focuses on two emotion categories: positive and negative, and also evaluates a three-class classification (positive, neutral, negative). For the binary setup, DEAP samples with valence scores above 4.5 are labeled as positive, and those below 4.5 as negative \cite{ni2021domain,salankar2021emotion,imtiaz2025enhanced}. SEED includes only positive and negative samples, excluding neutral ones. For DREAMER, samples with valence scores above 3 are labeled as positive, and those below 3 as negative \cite{zhang2024tpro,cheng2020emotion}. For multi-class experiments, in DEAP, valence scores from 1--4 are labeled negative, 4--6 neutral, and 6--9 positive \cite{tripathi2017using,nematollahi2017recognition}. SEED includes all three categories, and in DREAMER, scores from 1--2 are negative, 3 is neutral, and 4--5 is positive.

All experiments are performed on a Linux platform with Python (version 3.10.12) and the PyTorch library (version 2.5.1+cu121), utilizing an NVIDIA Tesla T4 GPU with 12GB of memory. The learning rate is configured at 0.0001, with a weight decay of 0.0005. We use a batch size of 64, employ the Rectified Linear Unit (ReLU) as the activation function, and optimize the model using the Adam optimizer. The hyperparameter $\alpha$ is set to 0.5, and the threshold $\tau$ is set to 0.9, consistent with our previous model \cite{imtiaz2025enhanced}.

\subsection{Results and Discussion}

To assess the model's performance, we conduct cross-dataset testing by training on one dataset and testing on the other, considering both directions. As no SF-UDA methods currently exist for EEG emotion recognition, we evaluate our approach by comparing it with four general state-of-the-art methods \cite{ragab2023source,zhao2023source,ahmed2023ssda,du2024generation}, known for their strong performance. We reimplemented these open-source methods and evaluated them on the same datasets used in our experiments to ensure a fair comparison.

Table \ref{table1} presents the overall accuracy comparisons across datasets for binary emotion (positive, negative) classification. Our method achieves the highest accuracy in both cases, 58.99\% for SEED $\rightarrow$ DEAP and 65.84\% for DEAP $\rightarrow$ SEED, surpassing the second-best methods by 5.44\% and 3.86\%, respectively. To further validate the significance of these performance differences, we conduct a paired-sample \textit{t-test} using \textit{p}-values. For DEAP $\rightarrow$ SEED, the accuracy differences are highly significant (**) when compared to all other approaches. For SEED $\rightarrow$ DEAP, the improvement is significant (*) when compared to the methods by Ragab et al. and Zhao et al., and highly significant compared to the remaining methods.

\begin{table}[!tb]
\scriptsize
\centering
\caption {Overall accuracy comparison of our proposed method with state-of-the-art SF-UDA methods for binary emotion (positive, negative) classification. Symbols represent accuracy differences (paired-sample \textit{t-test}: $\sim$ \textit{nonsignificant}, \textit{*p} $<$ 0.05, \textit{**p} $<$ 0.01).}
\begin{tabular}
{ p{0.32\linewidth} p{0.22\linewidth} p{0.17\linewidth}}
\hline
  & \multicolumn{2}{c}{Accuracy (\%)}\\
\cline{2-3}

 & SEED$\rightarrow$DEAP & DEAP$\rightarrow$SEED  \\

\hline
Ragab et al. \cite{ragab2023source}  & 53.55* & 59.57**\\
Zhao et al. \cite{zhao2023source} & 53.17* & 61.98**\\
Ahmed et al. \cite{ahmed2023ssda}  & 51.71** & 61.33**\\
Du et al. \cite{du2024generation}  & 48.74** & 56.45**\\
\textbf{Proposed method} & \textbf{58.99} & \textbf{65.84} \\
\hline
\end{tabular}
\label{table1}
\end{table}

To assess multi-class emotion recognition performance, we conduct additional experiments for three-class classification (positive, neutral, negative). As shown in Table \ref{table1_b}, our method again outperforms all other methods, achieving accuracies of 51.50\% for SEED $\rightarrow$ DEAP and 57.37\% for DEAP $\rightarrow$ SEED. The improvements are statistically highly significant in all cases, except for comparisons with Zhao et al., where they remain statistically significant.

\begin{table}[!tb]
\scriptsize
\centering
\caption {Overall accuracy comparison of our proposed method with state-of-the-art SF-UDA methods for three-class emotion (positive, neutral, negative) classification. Symbols represent accuracy differences (paired-sample \textit{t-test}: $\sim$ \textit{nonsignificant}, \textit{*p} $<$ 0.05, \textit{**p} $<$ 0.01).}
\begin{tabular}
{ p{0.32\linewidth} p{0.22\linewidth} p{0.17\linewidth}}
\hline
  & \multicolumn{2}{c}{Accuracy (\%)}\\
\cline{2-3}

 & SEED$\rightarrow$DEAP & DEAP$\rightarrow$SEED  \\

\hline
Ragab et al. \cite{ragab2023source}  & 45.97** & 51.82**\\
Zhao et al. \cite{zhao2023source} & 47.45* & 53.60*\\
Ahmed et al. \cite{ahmed2023ssda}  & 44.08** & 51.29**\\
Du et al. \cite{du2024generation}  & 39.93** & 47.35**\\
\textbf{Proposed method} & \textbf{51.50} & \textbf{57.37} \\
\hline
\end{tabular}
\label{table1_b}
\end{table}

To evaluate the robustness of our model in diverse laboratory settings, we further conduct experiments on the DREAMER dataset. Table \ref{table1_c} reports the performance of both binary (positive, negative) and multi-class (positive, neutral, negative) emotion recognition when the model is trained on either SEED or DEAP. Our method consistently outperforms all other approaches across both emotion settings, achieving accuracies of 67.08\% and 58.87\% for binary emotion recognition, and 59.44\% and 53.83\% for three-class emotion recognition when trained on SEED and DEAP, respectively. These improvements are statistically significant compared to all other approaches. Notably, the model performs slightly better when trained on SEED than on DEAP.

\begin{table}[!tb]
\scriptsize
\centering
\caption {Overall accuracy comparison of our proposed method with state-of-the-art SF-UDA methods for binary (positive, negative) and three-class emotion (positive, neutral, negative) classification, where the target dataset is DREAMER. Symbols represent accuracy differences (paired-sample \textit{t-test}: $\sim$ \textit{nonsignificant}, \textit{*p} $<$ 0.05, \textit{**p} $<$ 0.01).}
\begin{tabular}
{ p{0.16\linewidth} p{0.13\linewidth} p{0.20\linewidth} p{0.13\linewidth} p{0.12\linewidth}}
\hline
  & \multicolumn{2}{c}{Binary Emotion} & \multicolumn{2}{c}{Three-Class Emotion} \\

\hline
  & \multicolumn{4}{c}{Accuracy (\%)}\\
\cline{2-5}

 & SEED$\rightarrow$DREAMER & DEAP$\rightarrow$DREAMER & SEED$\rightarrow$DREAMER & DEAP$\rightarrow$DREAMER  \\

\hline
Ragab et al. \cite{ragab2023source}  & 58.26** & 54.90* &  49.02** & 49.15*\\
Zhao et al. \cite{zhao2023source} & 60.29** & 53.75** & 55.96* & 50.80*\\
Ahmed et al. \cite{ahmed2023ssda}  & 51.47** & 44.08** &  46.39** & 37.28**\\
Du et al. \cite{du2024generation}  & 55.65** & 54.46* & 48.65** & 49.00*\\
\textbf{Proposed method} & \textbf{67.08} & \textbf{58.87} & \textbf{59.44} & \textbf{53.83} \\
\hline
\end{tabular}
\label{table1_c}
\end{table}

\begin{figure}[!tb]
     \centering
    \begin{tabular}{cc}
\subfloat[]{\includegraphics[width=0.47\linewidth]{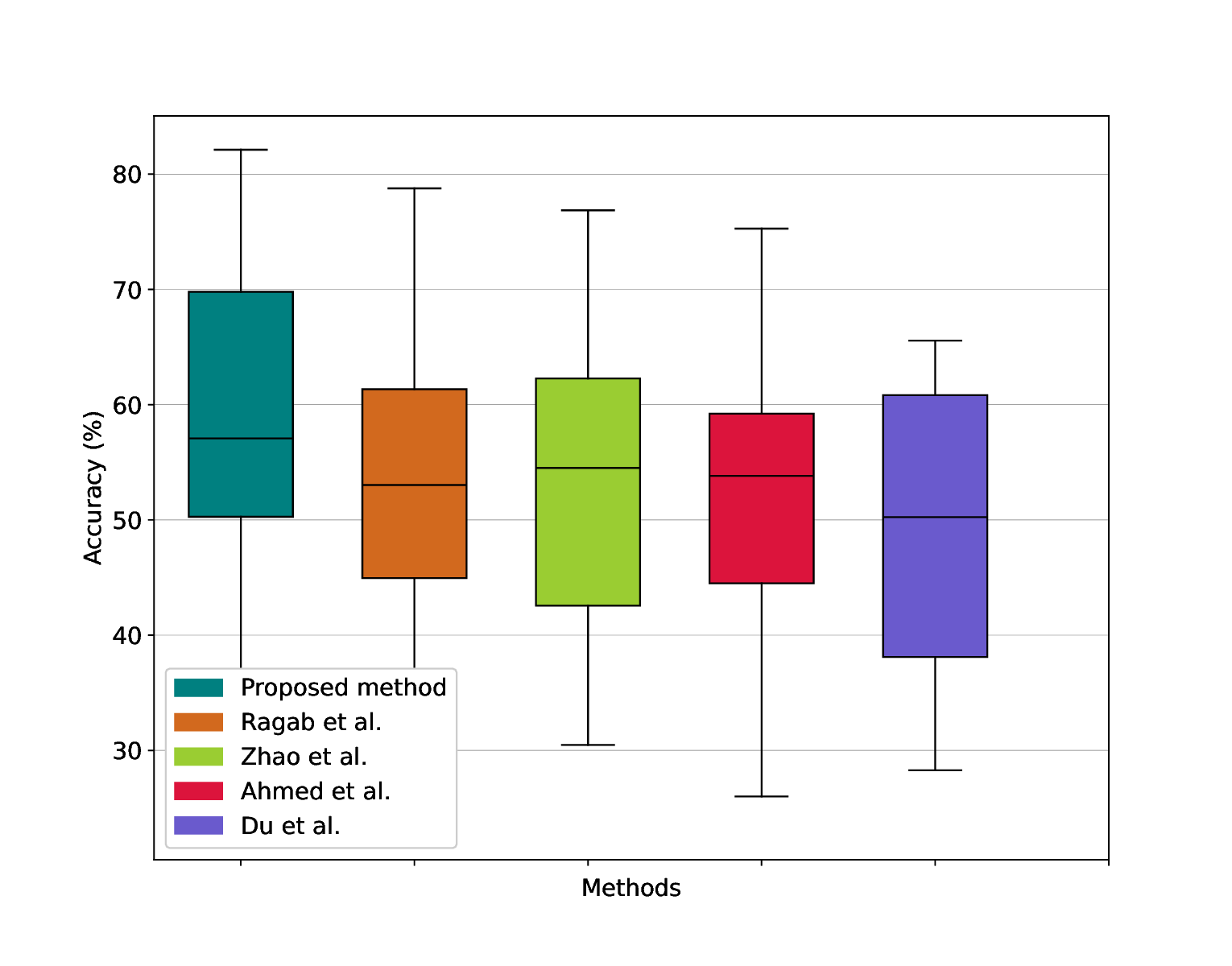}}
\subfloat[]{\includegraphics[width=0.47\linewidth]{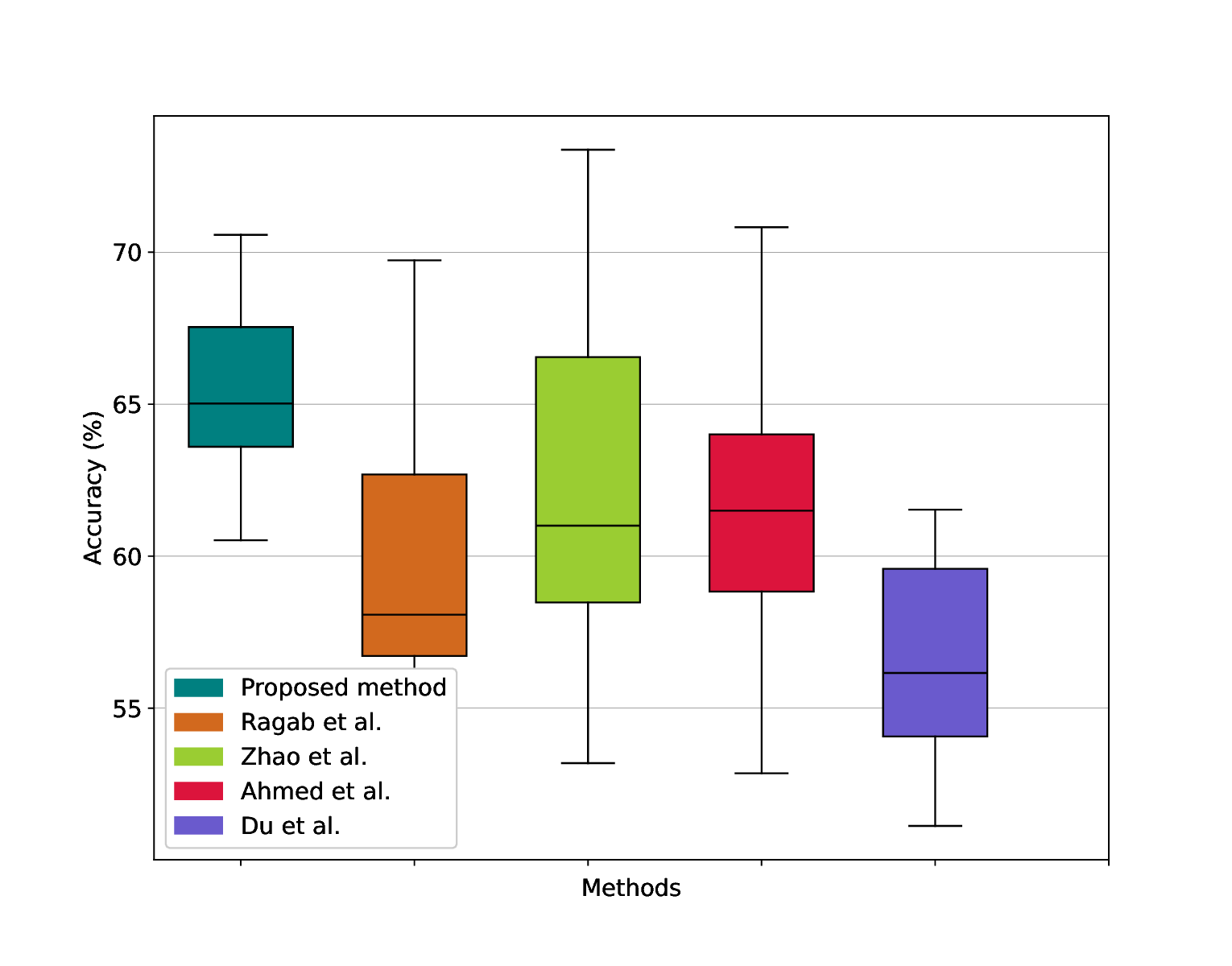}}
\end{tabular}

        \caption{Boxplots depicting the distribution of emotion recognition accuracies on the test dataset across subjects for our proposed method and other SF-UDA methods: (a) SEED $\rightarrow$ DEAP, (b) DEAP $\rightarrow$ SEED. Our method achieves the highest median accuracy in both cross-dataset scenarios and shows relatively low variation across subjects.}
        \label{fig4}
\end{figure}

Fig. \ref{fig4} displays boxplots of the accuracy distributions across subjects for our method and other SF-UDA approaches. Our method achieves the highest median accuracy in both scenarios. For DEAP $\rightarrow$ SEED, it attains a median accuracy of 65.02\%, with less variation compared to the others. In the SEED $\rightarrow$ DEAP scenario, although the methods by Ragab et al. and Ahmed et al. exhibit slightly lower variability, their accuracies are significantly lower, whereas our method achieves the highest median accuracy of 57.07\%.

Table \ref{table2} shows that our method significantly outperforms all other approaches in sensitivity, positive predictive value (PPV), and F1 score. On SEED, it achieves strong performance in detecting both positive and negative emotions, with F1 scores of 70.34\% and 59.73\%, respectively. On DEAP, all methods struggle with negative emotion detection. Although Zhao et al.'s method achieves a slightly higher F1 score for negative emotions (0.61\% higher than our method), our approach remains superior for positive emotions. The lower performance on DEAP may stem from its class imbalance (36.87\% negative vs. 63.13\% positive), whereas SEED has a more balanced distribution (50.36\% negative vs. 49.64\% positive).

\begin{table}[!tb]
\scriptsize
\centering
\caption {Comparison of the performance of our proposed method with other SF-UDA methods for detecting positive and negative emotions.}
\begin{tabular}
{p{0.16\linewidth} p{0.05\linewidth} p{0.04\linewidth}  p{0.04\linewidth} p{0.05\linewidth} p{0.04\linewidth} p{0.04\linewidth} p{0.05\linewidth} p{0.04\linewidth} p{0.04\linewidth} p{0.05\linewidth} p{0.04\linewidth} p{0.04\linewidth}  }
\hline
 &  \multicolumn{6}{c}{SEED$\rightarrow$DEAP}  & \multicolumn{6}{c}{DEAP$\rightarrow$SEED}\\
\cline{2-13}

 &  & Negative & & & Positive & & & Negative & & & Positive & \\
\cline{2-13}
& PPV (\%) & Se (\%) & F1 (\%) & PPV (\%) & Se (\%) & F1 (\%) & PPV (\%) & Se (\%) & F1 (\%) & PPV (\%) & Se (\%) & F1 (\%)\\
\hline
Ragab et al. \cite{ragab2023source} & 33.38 & 26.01 & 29.24 & 61.69 & 69.66 & 65.43 & 62.15 & 48.87 & 54.72 & 57.90 & 70.26 & 63.48\\

Zhao et al. \cite{zhao2023source} & 35.31 & 32.36 & \textbf{33.77} & 62.30 & 65.34 & 63.78 & 64.93 & 52.04 & 57.77 & 60.01 & 71.91 & 65.42\\

Ahmed et al. \cite{ahmed2023ssda} & 32.67 & 29.11 & 30.79 & 61.04 & 64.93 & 62.92 & 66.44 & 45.73 & 54.17 & 58.65 & 76.91 & 66.55\\

Du et al. \cite{du2024generation}  & 26.72 & 22.35 & 24.34 & 58.57 & 64.17 & 61.24 & 59.00 & 42.18 & 49.19 & 55.03 & 70.70 & 61.89\\

\textbf{Proposed method} & 41.59 & 27.57 & 33.16 & 64.63 & 77.37 & \textbf{70.43} & 72.70 & 50.69 & \textbf{59.73} & 62.17 & 80.97 & \textbf{70.34}\\
\hline
\end{tabular}
\label{table2}
\end{table}

\subsubsection{Experiments with different learning settings}
Table \ref{table3_a} presents the performance comparison of our proposed SF-UDA model against various learning settings, categorized based on source data access and target label availability. In all cases, we begin with a model pretrained on the source data. For the Supervised setting, we fine-tune this model using all available labeled target data. In the Semi-supervised setting, we use 10\% of the target labels and incorporate a supervised classification loss alongside our proposed loss functions. The UDA setting refers to our previous method \cite{imtiaz2025enhanced}, which performs domain adaptation using access to both labeled source and unlabeled target data. The Baseline (no adaptation) is derived from our SF-UDA model by removing the target adaptation stage entirely and directly evaluating the pretrained source model on the target domain.

\begin{table}[!tb]
\scriptsize
\centering
\caption {Comparison of overall accuracy under different learning settings.}
\begin{tabular}
{ p{0.275\linewidth} p{0.08\linewidth} p{0.14\linewidth} p{0.15\linewidth} p{0.14\linewidth}}
\hline
  & & & \multicolumn{2}{c}{Accuracy (\%)}\\
\cline{4-5}
Method & Source \: \: Access & Target Labels Used & SEED$\rightarrow$DEAP & DEAP$\rightarrow$SEED  \\
\hline
Supervised & \ding{55} & \ding{51} All & 65.09 & 70.87\\
Semi-supervised & \ding{55} & \ding{51} Partial & 61.52 & 66.93\\
UDA \cite{imtiaz2025enhanced} & \ding{51} & \ding{55} & 59.68 & 67.44\\
Baseline (no adaptation)& \ding{55} & \ding{55} & 48.66 & 54.24\\
\textbf{SF-UDA (proposed)} & \ding{55} & \ding{55} & 58.99 & 65.84\\
\hline
\end{tabular}
\label{table3_a}
\end{table}

The supervised model yields the highest performance, serving as an upper bound. However, its effectiveness is limited by substantial distributional differences between the source and target domains, as well as the pretrained model's inherent bias toward source-specific features. The absence of source data and target labels leads to a substantial drop in accuracy. Nevertheless, our proposed SF-UDA model significantly outperforms the baseline despite operating under the same constraints (no access to source data or target labels), demonstrating its ability to effectively reduce domain discrepancy in a fully source-free setting.

\subsubsection{Ablation study}

\begin{table}[!tb]
\scriptsize
\centering
\caption {Comparison of overall accuracy across both datasets in the ablation study.}
\begin{tabular}
{p{0.28\linewidth} p{0.13\linewidth} p{0.15\linewidth}  p{0.13\linewidth} p{0.05\linewidth} }
\hline
&  \multicolumn{4}{c}{Accuracy (\%)} \\
\cline{2-5}
 & SEED$\rightarrow$DEAP &  & DEAP$\rightarrow$SEED &\\
\cline{2-5}
& PSD & DE & PSD  & DE \\
\hline
Model A & 48.66 & 48.03 & 54.24 & 54.07\\

Model B & 54.02 & 53.19 & 60.82 & 60.95 \\

Model C & 55.50 & 55.42 & 62.64 & 61.41 \\

\textbf{Proposed method} & \textbf{58.99} & 58.47 & \textbf{65.84} & 64.76 \\
\hline
\end{tabular}
\label{table3}
\end{table}

\begin{table}[!tb]
\scriptsize
\centering
\caption {Ablation on Model B losses.}
\begin{tabular}
{p{0.28\linewidth} p{0.22\linewidth} p{0.18\linewidth}  }
\hline
&  \multicolumn{2}{c}{Accuracy (\%)} \\
\cline{2-3}
 Variation of Model B & SEED$\rightarrow$DEAP &  DEAP$\rightarrow$SEED \\

\hline
Full & 54.02 & 60.82 \\

Without $L_{plal}$ & 51.18 & 57.61  \\

Without $L_{ccdl}$ & 52.33 & 59.45  \\

\hline
\end{tabular}
\label{table3_b}
\end{table}

We conduct an ablation study to evaluate the contribution of each component in our proposed method. This involves systematically removing key components, which are the primary contributions of this study, and assessing their impact on the model's performance. Three models are created by excluding one component at a time while keeping all other aspects unchanged. Model A eliminates the entire target adaptation stage. Model B removes the Localized Consistency Learning (LCL) strategy from the target adaptation stage. Model C is formed by excluding the Dual-Loss Adaptive Regularization (DLAR) technique from the target adaptation stage while retaining all other components.

\begin{figure}[!tb]
     \centering
    \begin{tabular}{cc}
\subfloat[]{\includegraphics[width=0.47\linewidth]{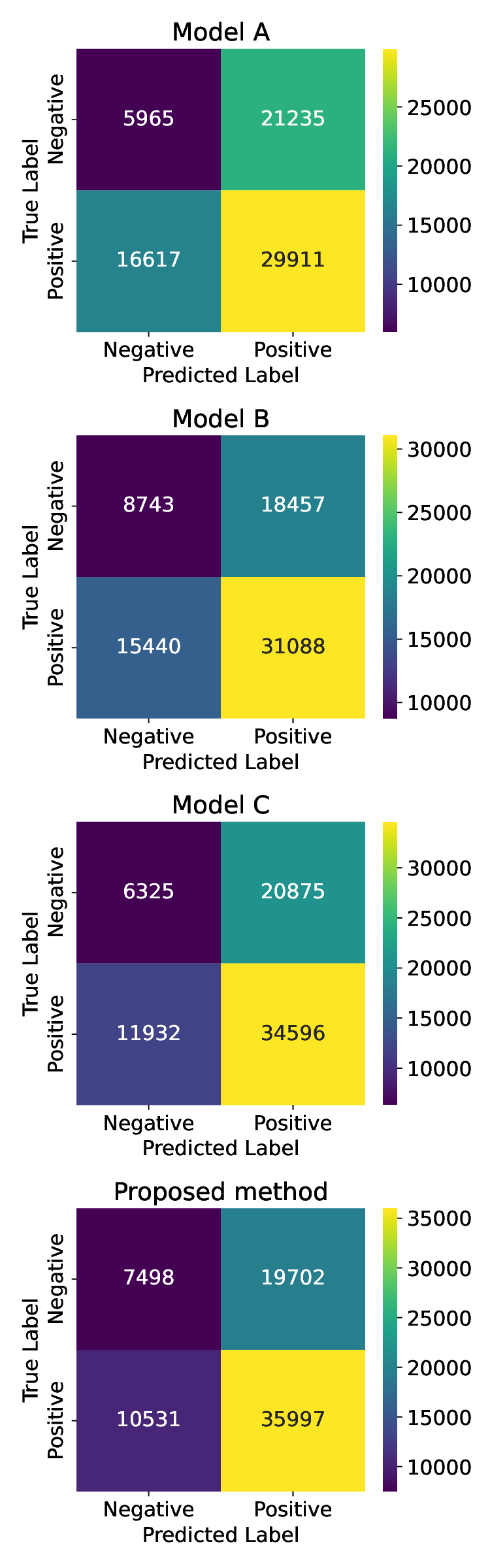}} \; \;
\subfloat[]{\includegraphics[width=0.47\linewidth]{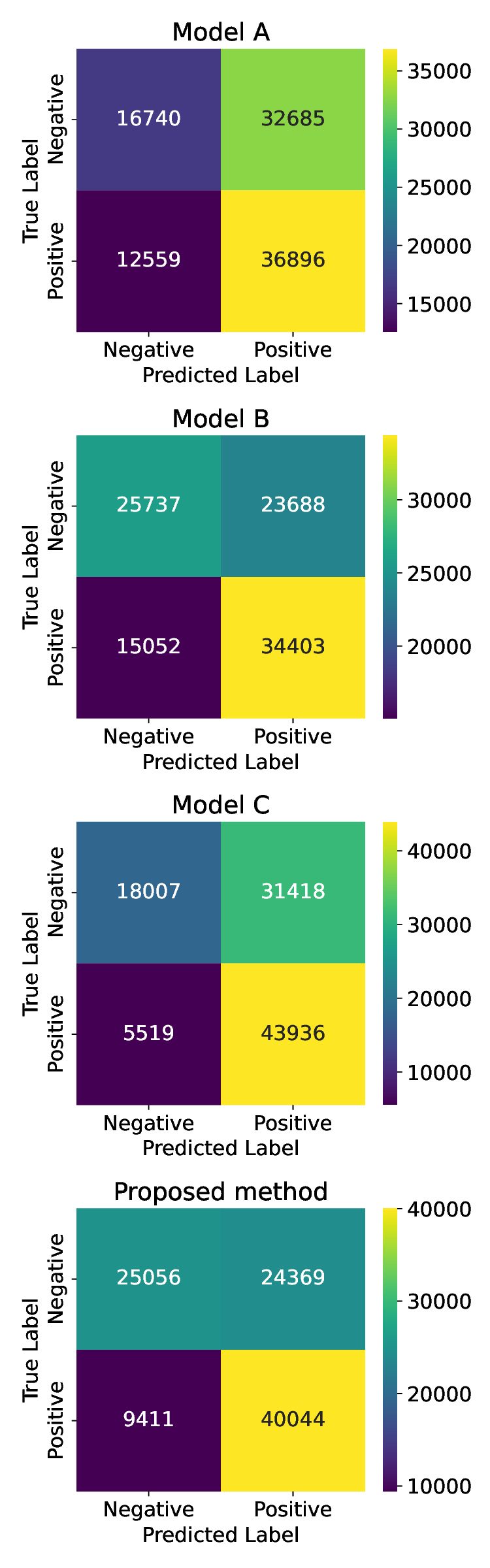}}
\end{tabular}

        \caption{Confusion matrices from the ablation study: (a) SEED $\rightarrow$ DEAP, (b) DEAP $\rightarrow$ SEED. The LCL component improves recognition of positive emotions, while the DLAR component enhances detection of negative emotions.}
        \label{fig5}
\end{figure}

Along with evaluating the models using Power Spectral Density (PSD), we also assess their performance with Differential Entropy (DE) features. Table \ref{table3} shows the performance accuracy of all models using both PSD and DE features. As shown in Table \ref{table3}, Table \ref{table3_b}, Table \ref{table4}, and Fig. \ref{fig5}, each component of the proposed model contributes to the overall performance to some extent. In general, PSD features lead to better performance, although DE performs slightly better for Model B in the DEAP $\rightarrow$ SEED scenario. PSD consistently yields strong results; therefore, we select it for our proposed approach and will use it for comparisons moving forward. Model A performs the worst, even falling below random probability for SEED $\rightarrow$ DEAP, due to the absence of any information from the target domain. Both the DLAR and LCL techniques play crucial roles in enhancing emotion classification performance (Table \ref{table3}). Excluding the LCL strategy (Model B) leads to a substantial drop in accuracy, with a reduction of 4.97\% for SEED $\rightarrow$ DEAP and 5.02\% for DEAP $\rightarrow$ SEED. Removing the DLAR technique (Model C) results in a decrease in overall accuracy of 3.49\% for SEED $\rightarrow$ DEAP and 3.20\% for DEAP $\rightarrow$ SEED. Table \ref{table3_b} further presents an ablation study on Model B using PSD features, where one of the two losses, $L_{plal}$ or $L_{ccdl}$, is removed at a time. Although both losses contribute to performance, $L_{plal}$ has a slightly greater impact.

The DLAR and LCL components significantly improve the model's performance in classifying both positive and negative emotions, as illustrated in Table \ref{table4} and Fig. \ref{fig5}. From the confusion matrices in Fig. \ref{fig5}, it is evident that LCL substantially improves the recognition of positive emotions across both datasets, while DLAR notably enhances the detection of negative emotions. According to Table \ref{table4}, DLAR substantially improves F1 scores for negative emotion detection, with gains of 10.07\% (SEED $\rightarrow$ DEAP) and 14.53\% (DEAP $\rightarrow$ SEED). LCL boosts positive emotion detection by 6.59\% and 8.41\% in the same directions. DLAR enhances the model's consistency and reliability, while LCL promotes local consistency among reliable neighbors. Together, they effectively improve the model's generalization to target data.

\begin{table}[!tb]
\scriptsize
\centering
\caption {Performance of the components in the proposed method for detecting positive and negative emotions.}
\begin{tabular}
{p{0.15\linewidth} p{0.05\linewidth} p{0.04\linewidth}  p{0.04\linewidth} p{0.05\linewidth} p{0.04\linewidth} p{0.04\linewidth} p{0.05\linewidth} p{0.04\linewidth} p{0.04\linewidth} p{0.05\linewidth} p{0.04\linewidth} p{0.04\linewidth} }
\hline
 &  \multicolumn{6}{c}{SEED$\rightarrow$DEAP}  & \multicolumn{6}{c}{DEAP$\rightarrow$SEED} \\
\cline{2-13}
 &  & Negative & & & Positive & & & Negative & & & Positive & \\
\cline{2-13}
& PPV (\%) & Se (\%) & F1 (\%) & PPV (\%) & Se (\%) & F1 (\%) & PPV (\%) & Se (\%) & F1 (\%) & PPV (\%) & Se (\%) & F1 (\%)\\
\hline
Model A & 26.41 & 21.93 & 23.96 & 58.48 & 64.29 & 61.25 & 57.14 & 33.87 & 42.53 & 53.03 & 74.61 & 62.00\\

Model B & 36.15 & 32.14 & \textbf{34.03} & 62.75 & 66.82 & 64.72 & 63.10 & 52.07 & 57.06 & 59.22 & 69.56 & 63.97\\

Model C & 34.64 & 23.25 & 27.82 & 62.37 & 74.36 & 67.84 & 76.54 & 36.43 & 49.36 & 58.31 & 88.84 & \textbf{70.41}\\

\textbf{Proposed method} & 41.59 & 27.57 & 33.16 & 64.63 & 77.37 & \textbf{70.43} & 72.70 & 50.69 & \textbf{59.73} & 62.17 & 80.97 & 70.34\\
\hline
\end{tabular}
\label{table4}
\end{table}

\begin{table}[!tb]
\scriptsize
\centering
\caption {Overall accuracy comparison for different neighbor selection scenarios.}
\begin{tabular}
{ p{0.10\linewidth} p{0.36\linewidth} p{0.19\linewidth} p{0.16\linewidth}}
\hline
   &  & \multicolumn{2}{c}{Accuracy (\%)}\\
\cline{3-4}
 \multicolumn{2}{c}{Neighbor Selection}  & SEED$\rightarrow$DEAP & DEAP$\rightarrow$SEED \\
\hline
$k$=5 & Features only & 58.14 & 64.96\\
 & Softmax outputs only & 56.52 &  62.47\\
 & Features and Softmax outputs & \textbf{58.99} & \textbf{65.84}\\
 $k$=10 & Features only & 57.56 & 64.28\\
 & Softmax outputs only& 55.43 & 63.05\\
 & Features and Softmax outputs & 57.40 & 65.72\\
 $k$=15 & Features only & 56.34 & 63.90\\
 & Softmax outputs only & 54.73 & 60.96\\
 & Features and Softmax outputs & 57.55 & 64.98 \\
\end{tabular}
\label{table5}
\end{table}

\begin{figure}[!tb]
     \centering
    \begin{tabular}{cc}
\subfloat[]{\includegraphics[width=0.47\linewidth]{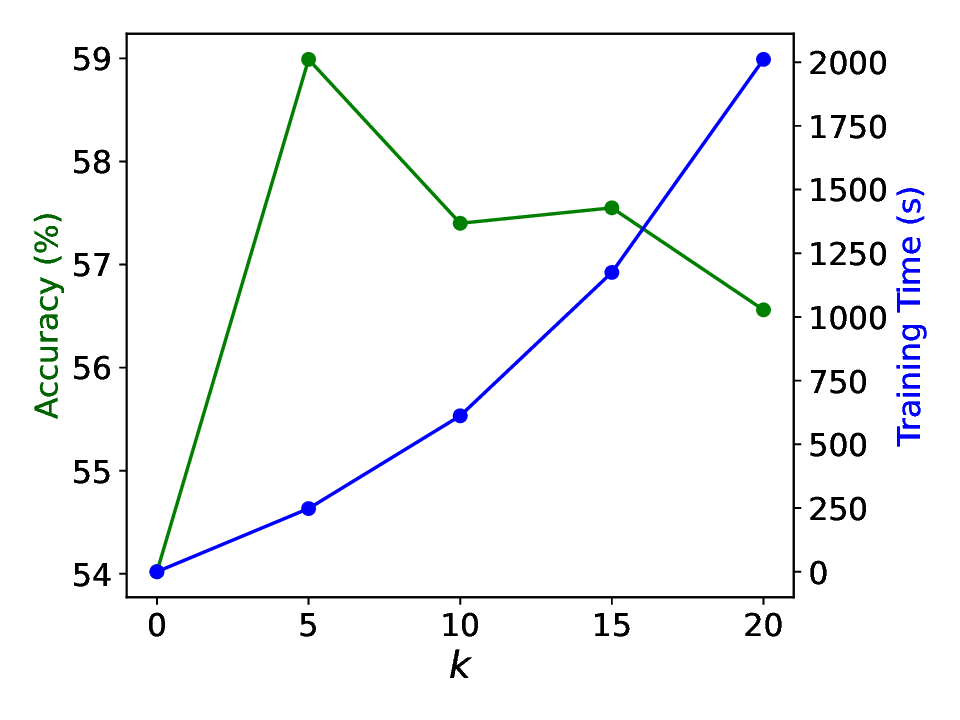}} \; \;
\subfloat[]{\includegraphics[width=0.47\linewidth]{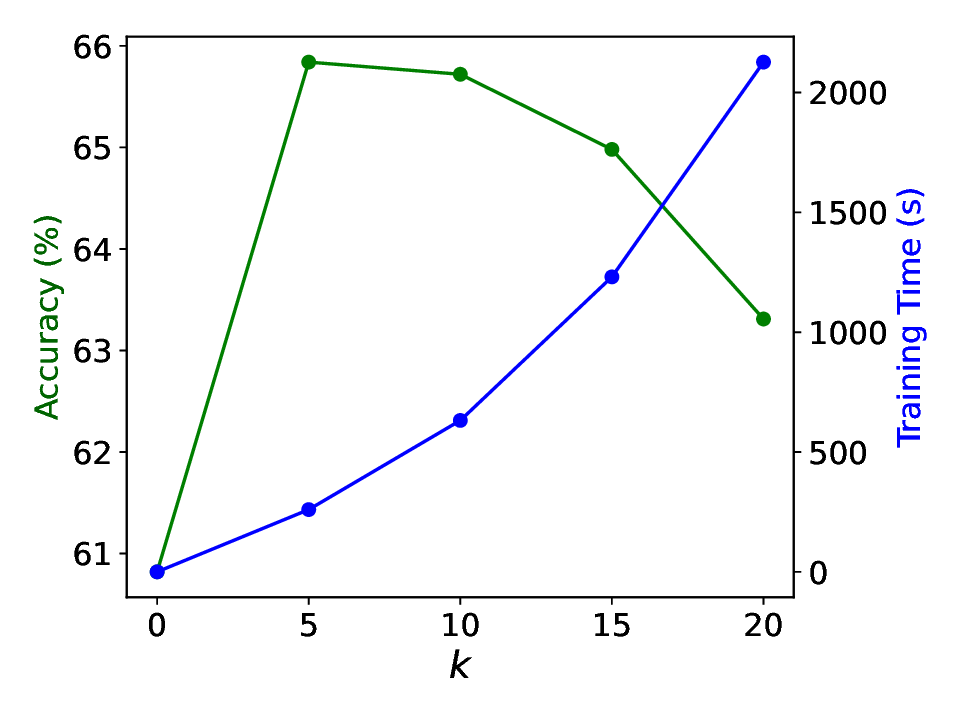}}
\end{tabular}

        \caption{Model performance for varying values of $k$ in the LCL strategy: (a) SEED $\rightarrow$ DEAP, (b) DEAP $\rightarrow$ SEED. The model achieves the highest accuracy at $k = 5$ in both settings.}
        \label{fig7}
\end{figure}

To evaluate the impact of different neighbor selection scenarios in the Localized Consistency Learning (LCL) strategy, we test three cases: using feature representations only, softmax probabilities only, and a combination of both. For each scenario, we vary the number of neighbors, denoted as $k$. Table \ref{table5} demonstrates that considering both feature representations and softmax probabilities consistently yields better results compared to considering either one alone. This approach helps identify more reliable neighbors while minimizing the influence of unreliable ones. Fig. \ref{fig7} illustrates the model's accuracy and training time (in seconds) for the LCL strategy as $k$ is varied. The model achieves its highest accuracy with $k$ = 5 for both SEED $\rightarrow$ DEAP and DEAP $\rightarrow$ SEED. Additionally, we observe that training time increases exponentially as $k$ increases. Based on these findings, we set $k$ = 5 for our proposed model.

\begin{table}[!tb]
\scriptsize
\centering
\caption {Ablation study on pseudo-label assignment strategies in the DLAR stage.}
\begin{tabular}
{ p{0.32\linewidth} p{0.19\linewidth} p{0.18\linewidth} p{0.15\linewidth}}
\hline
  & & \multicolumn{2}{c}{Accuracy (\%)}\\
\cline{3-4}
Pseudo-label Type & Temperature (T) & SEED$\rightarrow$DEAP & DEAP$\rightarrow$SEED  \\
\hline
Our approach & --- & \textbf{58.99} & \textbf{65.84}\\
Softmax over distances & 1 & 58.52 & 63.80\\
Temperature-scaled softmax & 2 & 58.21 & 64.19\\
Temperature-scaled softmax & 3 & 56.84 & 62.95\\
\hline
\end{tabular}
\label{table6}
\end{table}

We perform an ablation study to evaluate different pseudo-label assignment strategies in the DLAR stage. Our approach assigns pseudo-labels based on the nearest centroid, and we compare it with soft pseudo-labels generated using a temperature-scaled softmax over centroid distances. While soft assignments provide smoother labels, our direct assignment yields better results in our experiments (Table \ref{table6}). This suggests that the model rapidly improves feature representations for the target domain during training, leading to better centroid separability within DLAR. Hard assignments enforce decisive class boundaries, stabilizing training by avoiding the spread of uncertain pseudo-labels. In contrast, soft labels diffuse the supervision signal across classes, which increases the model's sensitivity to noise and temporal variability in EEG features.

\subsubsection{Evolution of cluster quality and pseudo-label accuracy in DLAR}
\begin{figure}[!tb]
     \centering
    \begin{tabular}{cc}
\subfloat[]{\includegraphics[width=0.47\linewidth]{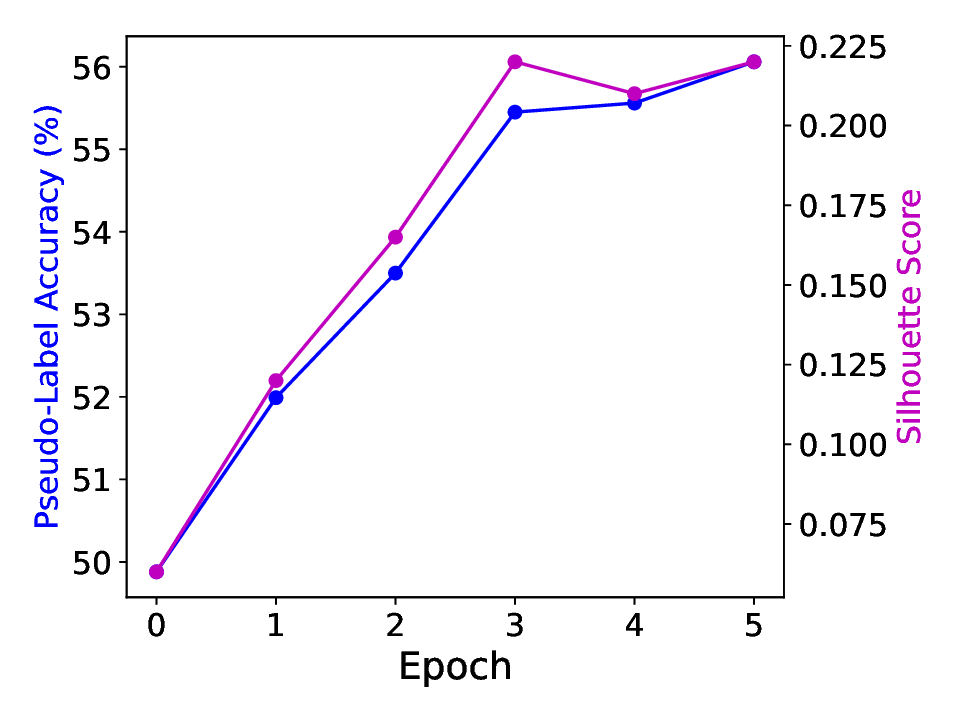}} \; \;
\subfloat[]{\includegraphics[width=0.47\linewidth]{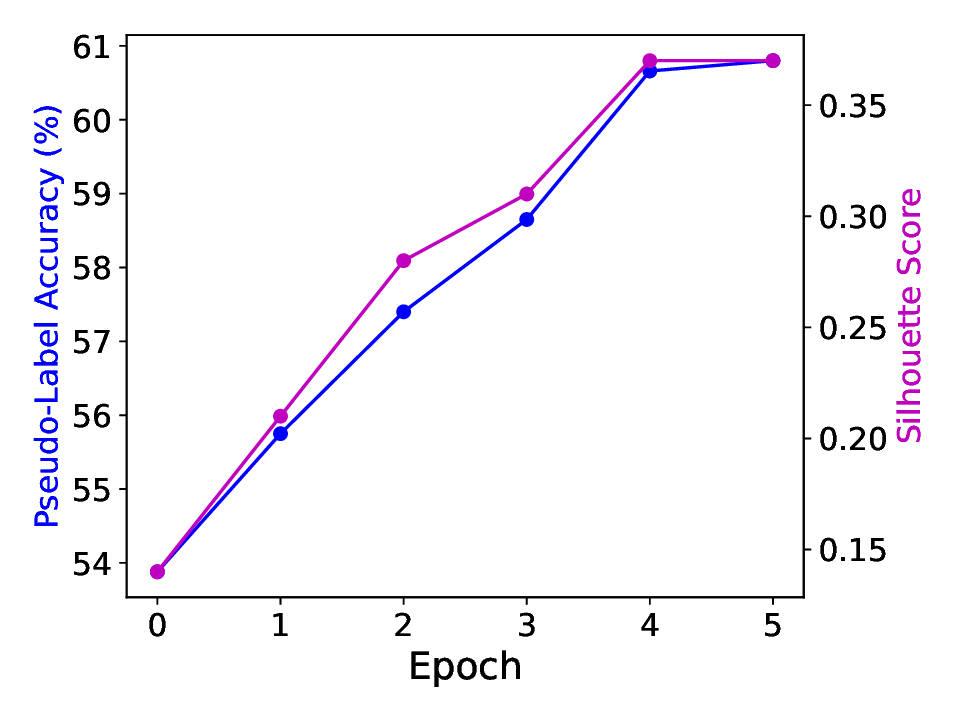}}
\end{tabular}

        \caption{Evolution of cluster quality (measured by Silhouette score) and pseudo-label accuracy during the DLAR stage: (a) SEED $\rightarrow$ DEAP, (b) DEAP $\rightarrow$ SEED. Both metrics steadily improve throughout training.}
        \label{fig7_b}
\end{figure}

To analyze the progression of cluster quality and pseudo-label correctness during the DLAR stage, we conduct additional experiments. We use a small held-out labeled set from the target domain to evaluate the accuracy of pseudo-labels generated in the DLAR stage against the true labels. Additionally, we quantify the evolution of cluster organization using the Silhouette score, which measures clustering quality by evaluating how well each sample fits within its assigned cluster relative to other clusters. Fig. \ref{fig7_b} shows the pseudo-label accuracy and cluster quality, measured by the Silhouette score, across training epochs. As the figure illustrates, both pseudo-label accuracy and cluster quality steadily improve as training progresses, indicating that the feature representations become increasingly discriminative during the DLAR stage.

\subsubsection{Runtime analysis of model stages}
The computational time (in seconds) for the various stages of our model is reported in Table \ref{table6_b}. The LCL stage is the most computationally intensive, as it involves more training epochs than the DLAR stage and requires identifying the nearest neighbors for each sample by considering both the feature representations and the softmax probabilities of all samples in the training batch.

\begin{table}[!tb]
\scriptsize
\centering
\caption {Computational time of different model stages.}
\begin{tabular}
{ p{0.25\linewidth} p{0.21\linewidth} p{0.16\linewidth}}
\hline
  & \multicolumn{2}{c}{Execution Time (s)}\\
\cline{2-3}
Stage & SEED$\rightarrow$DEAP & DEAP$\rightarrow$SEED  \\
\hline
DLAR & 100.18 & 117.66\\
LCL&  247.95 & 259.79\\
Inference &  48.91 & 56.06\\
\hline
\end{tabular}
\label{table6_b}
\end{table}

\subsubsection{Failure analysis}
\begin{figure}[!tb]
     \centering
    \begin{tabular}{cc}
\subfloat[]{\includegraphics[width=0.47\linewidth]{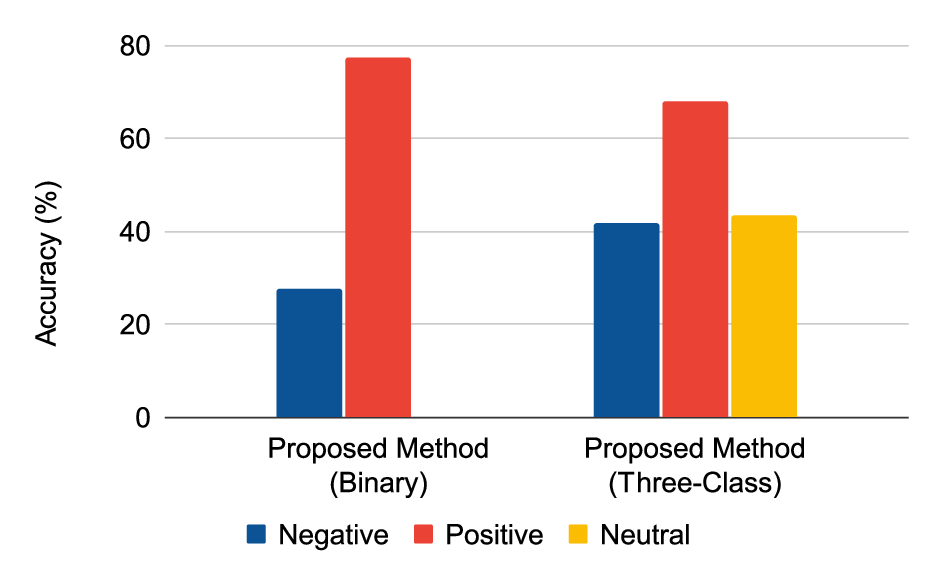}} \; \;
\subfloat[]{\includegraphics[width=0.47\linewidth]{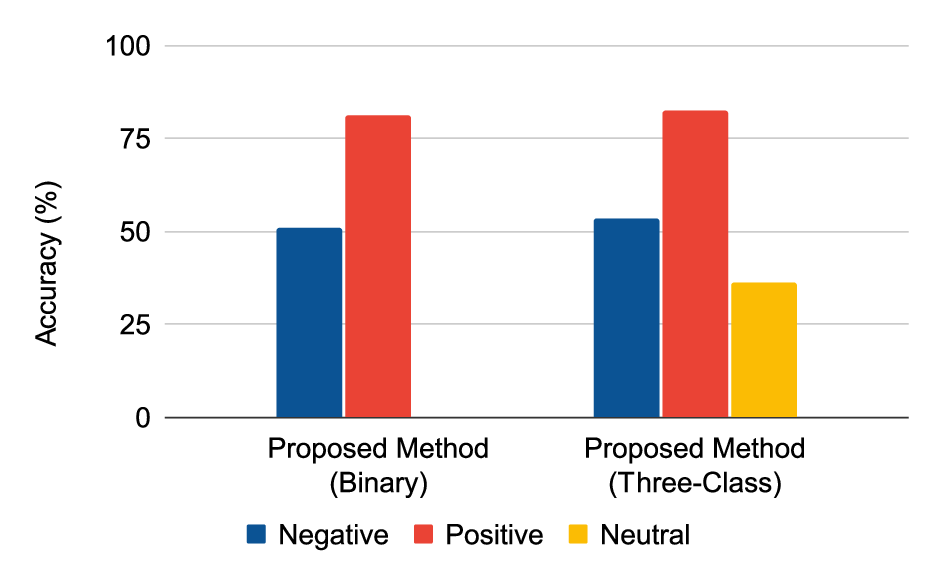}}
\end{tabular}

        \caption{Class-wise accuracy of the proposed method for binary (positive, negative) and three-class (positive, neutral, negative) emotion classification: (a) SEED $\rightarrow$ DEAP, (b) DEAP $\rightarrow$ SEED. Lower negative-class accuracy is due to class imbalance, while reduced neutral-class accuracy stems from class ambiguity.}
        \label{fig7_c}
\end{figure}

Although our model performs significantly better than existing methods, its performance degrades in certain scenarios. For instance, as shown in Fig. \ref{fig7_c}, lower accuracy is observed for negative emotion classification on the DEAP dataset in the binary setting, primarily due to class imbalance (36.87\% negative vs. 63.13\% positive samples). In contrast, in the three-class setting, the performance on the negative class improves as the class distribution becomes more balanced (27.81\% negative, 30.08\% neutral, and 42.11\% positive). Despite being relatively balanced, the neutral class exhibits lower accuracy due to its lack of distinctive features and significant overlap with both positive and negative emotions.

Overall, Tables \ref{table1}, \ref{table1_b}, \ref{table3}, and \ref{table4} indicate that the model achieves comparatively lower performance on DEAP than on SEED. One key reason is that DEAP exhibits greater subject diversity than SEED. While all SEED participants are of Chinese ethnicity, DEAP includes participants from across Europe, introducing higher inter-subject variability. In addition, SEED provides more consistent EEG recordings with a larger number of samples per subject. As illustrated in Fig. \ref{fig4}, the variation in classification accuracy across subjects is notably higher for DEAP than for SEED. This can be partly explained by the larger number of participants in DEAP (32) compared to SEED (15), where substantial inter-individual differences play a critical role in performance degradation.

Finally, differences in stimulus types also influence the results. SEED and DREAMER employ film clips as emotional stimuli, whereas DEAP uses music videos. Consequently, performance transfer between SEED and DREAMER is higher (Table \ref{table1_c}) than that between SEED and DEAP (Tables \ref{table1} and \ref{table1_b}), as well as between DEAP and DREAMER (Table \ref{table1_c}).

\subsubsection{EEG feature importance analysis}
\begin{figure}[!tb]
     \centering
    \begin{tabular}{cc}
\subfloat[]{\includegraphics[width=0.47\linewidth]{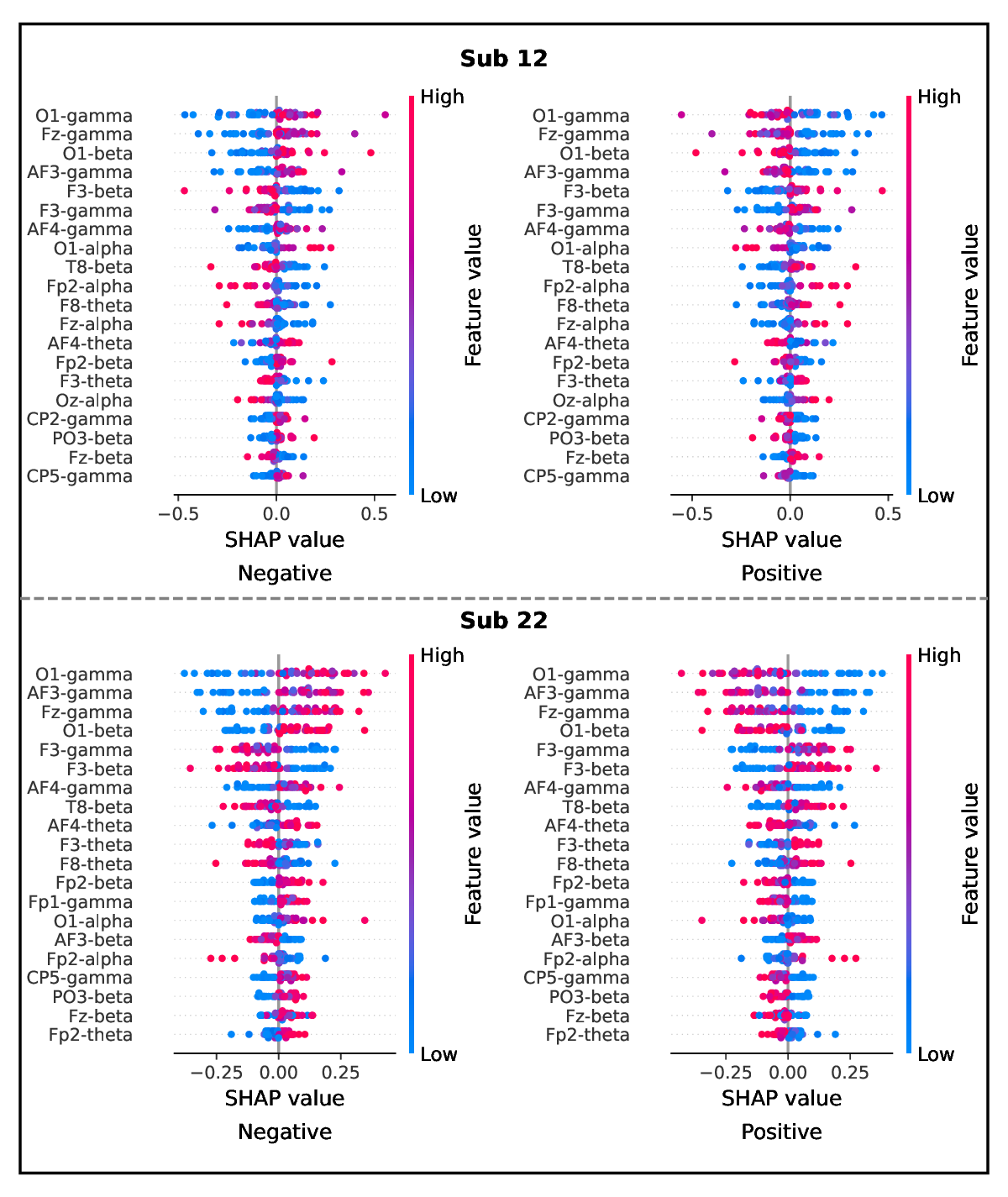}} \;
\subfloat[]{\includegraphics[width=0.47\linewidth]{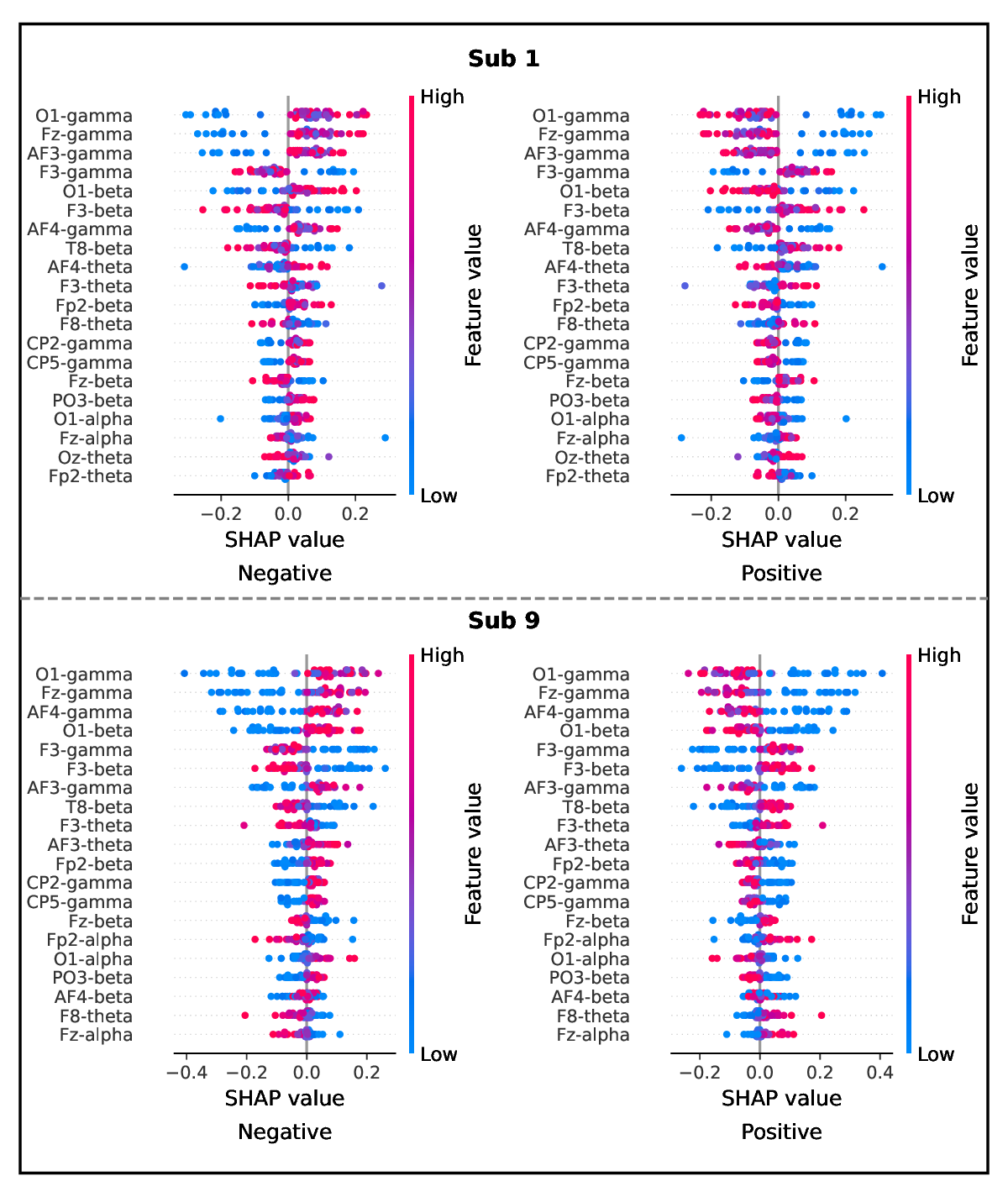}}
\end{tabular}

        \caption{Distributions of SHAP values for (a) SEED $\rightarrow$ DEAP and (b) DEAP $\rightarrow$ SEED. For the DEAP dataset, Subjects 12 and 22 are considered, while for the SEED dataset, Subjects 1 and 9 are selected based on prediction accuracies close to their respective dataset averages. SHAP values are computed for both positive and negative emotions and visualized for the top 20 most important EEG features. Each feature is labeled using the format Channel--Band (e.g., Fz--gamma). Each dot represents one feature value from a test sample, with the x-axis indicating its SHAP value, which reflects the direction and magnitude of the feature's impact on predictions. A wider spread or denser distribution along the x-axis indicates a stronger influence. Dot colors correspond to the actual feature values, ranging from blue (low) to red (high). EEG features from frontal, prefrontal, occipital, and temporal regions, particularly in high-frequency bands, have a strong influence on emotion prediction.}
        \label{fig8}
\end{figure}

To understand the model's decision-making process and identify the most influential EEG features for emotion recognition, we analyze SHapley Additive exPlanations (SHAP) values \cite{lundberg2017unified}. Fig. \ref{fig8} shows the SHAP value distributions for representative subjects in the SEED $\rightarrow$ DEAP and DEAP $\rightarrow$ SEED settings. SHAP values are computed for positive and negative emotions, and among all channel--band pairs (five frequency bands across 32 EEG channels), the top 20 features are visualized based on average absolute SHAP values. The analysis identifies O1, Fz, AF3, AF4, F3, T8, Fp2, and F8 as the most pivotal EEG channels, reflecting the involvement of the occipital, frontal, prefrontal, and temporal regions, areas well established in emotional processing and cognitive function \cite{lindquist2012brain,nakajima2021preserving}. Among the frequency bands, gamma, beta, and alpha consistently emerge as key contributors, highlighting the importance of high-frequency oscillations in cognition and emotional regulation for driving the model's classification performance. These findings align with neuroscience evidence emphasizing the critical role of these bands in modulating emotional and cognitive processes \cite{mouri2023identifying,park2011emotion}.

\subsubsection{Robustness evaluation under noisy EEG conditions}

\begin{table}[!tb]
\scriptsize
\centering
\caption {Evaluation of model robustness under synthetic EEG noise. $\sigma$ denotes the standard deviation of Gaussian noise. $\Delta$ Acc indicates the accuracy drop relative to the clean input.}
\begin{tabular}
{ p{0.19\linewidth} p{0.22\linewidth} p{0.05\linewidth} p{0.08\linewidth} p{0.05\linewidth} p{0.08\linewidth}}
\hline
  & &  \multicolumn{2}{c}{ SEED$\rightarrow$DEAP} & \multicolumn{2}{c}{ DEAP$\rightarrow$SEED}\\
\cline{3-6}
Noise Type & Parameters &  Acc (\%) & $\Delta$ Acc & Acc (\%) & $\Delta$ Acc\\
\hline
Clean (no noise) & --- & 58.99 & --- & 65.84 & --- \\
Gaussian noise & $\sigma = 0.1$ & 58.99 & 0.0 & 65.88 & +0.04 \\
Gaussian noise & $\sigma = 0.5$ & 58.97 & -0.02 & 65.84 & 0.0 \\
Gaussian noise & $\sigma = 0.9$ & 58.95 & -0.04 & 65.79 & -0.05 \\
Powerline noise & 50 Hz sinusoidal interference & 58.99 & 0.0 & 65.80 & -0.04 \\
Baseline drift & Low-freq sinusoid (0.5 Hz)  & 59.00 & +0.01 & 65.83 & -0.01 \\
\end{tabular}
\label{table8}
\end{table}

EEG data are often contaminated by various types of noise. To evaluate the robustness of our model, we test it under several synthetic EEG noise conditions. Specifically, we add common noise types to the original clean signals: Gaussian noise at three different standard deviations ($\sigma = 0.1, 0.5, 0.9$), powerline noise simulating 50 Hz interference, and baseline drift modeled as a low-frequency sinusoid (0.5 Hz), then evaluate the model. As summarized in Table \ref{table8}, our model exhibits stable performance across all noise types and levels, with only minimal accuracy degradation and occasional slight improvements. These results demonstrate the model's strong resilience to typical EEG signal corruptions encountered in real-world scenarios.

Our findings reveal that the proposed method delivers robust performance across different datasets. Despite encountering substantial distributional differences between the training and testing data and the absence of source domain data, it consistently outperforms other state-of-the-art approaches.

\section{Conclusion}
\label{conclusion}
This study introduces a novel source-free unsupervised domain adaptation (SF-UDA) method for EEG-based emotion recognition, addressing critical challenges such as domain discrepancies, limited labeled data, and the absence of source data due to privacy and computational constraints. To the best of our knowledge, this is the first application of SF-UDA in this domain. The proposed approach incorporates Dual-Loss Adaptive Regularization (DLAR) to enhance model consistency and reliability through classifier agreement and alignment with the underlying cluster distributions. Additionally, Localized Consistency Learning (LCL) promotes consistent predictions among reliable neighbors, supported by a mechanism to identify trustworthy neighbors based on feature values and softmax probabilities. Cross-dataset results demonstrate that the method significantly outperforms state-of-the-art approaches. Its simplicity and adaptability make it well-suited for resource-constrained environments. This work has substantial industrial potential, enabling empathetic human-computer interactions and reliable mood disorder management in healthcare, while safeguarding data privacy and intellectual property. In future work, we aim to evaluate the generalizability of our model on a wider range of datasets, refine the pseudolabeling process, and extend it to support multiple target domains. We also plan to adapt the model for continuous emotion labeling based on valence and arousal dimensions, enabling dynamic emotion recognition over time. Furthermore, we intend to investigate alternative uncertainty estimation methods, such as predictive variance and mutual information, to more effectively identify low-confidence predictions for test-time augmentation.

\section*{Acknowledgments}
The authors gratefully acknowledge the financial support provided by the Natural Sciences and Engineering Research Council of Canada (NSERC) and the New Frontiers in Research Fund (NFRF).

\end{document}